\documentclass[fleqn,usenatbib]{mnras}

\usepackage{newtxtext,newtxmath}

\usepackage[T1]{fontenc}

\DeclareRobustCommand{\VAN}[3]{#2}
\let\VANthebibliography\thebibliography
\def\thebibliography{\DeclareRobustCommand{\VAN}[3]{##3}\VANthebibliography}

\usepackage{graphicx}	
\usepackage{amsmath}	
\usepackage{upgreek}
\usepackage[export]{adjustbox}
\usepackage{booktabs}
\usepackage{hyperref}
\usepackage{subcaption}
\usepackage{url}
\usepackage[dvipsnames]{xcolor}

\newcommand{\dsigma}{\ensuremath{\Delta\sigma}}
\newcommand{\dv}{\ensuremath{\Delta v}}

\newcommand{\ewha}{\ensuremath{\rm EW(H{\upalpha})}}

\newcommand{\forb}[2]{[#1\,\textsc{#2}]}
\newcommand{\ha}{\ensuremath{{\rm H}\upalpha}}
\newcommand{\hb}{\ensuremath{{\rm H}\upbeta}}

\newcommand{\hi}{H\,\textsc{i}}
\newcommand{\hii}{H\,\textsc{ii}}

\newcommand{\kms}{\ensuremath{\rm \,km\,s^{-1}}}
\newcommand{\kpc}{\ensuremath{{\rm kpc}}}
\newcommand{\lzifu}{\textsc{LZIFU}}
\newcommand{\lzcomp}{\textsc{LZComp}}

\newcommand{\mstar}{\ensuremath{M_*}}
\newcommand{\msol}{\ensuremath{\rm \,M_\odot}}

\newcommand{\re}{\ensuremath{R_{\rm e}}}
\newcommand{\sfrdens}[1]{\ensuremath{\Sigma_{{\rm SFR},\,#1}}}
\newcommand{\sfr}[1]{\ensuremath{{\rm SFR}_{#1}}}
\newcommand{\sigmagas}{\ensuremath{\sigma_{\rm gas}}}

\newcommand{\sigmainst}{\ensuremath{\sigma_{\rm inst}}}
\newcommand{\sigmanarrow}{\ensuremath{\sigma_{\rm narrow}}}
\newcommand{\sigmabroad}{\ensuremath{\sigma_{\rm broad}}}
\newcommand{\vgas}{\ensuremath{v_{\rm gas}}}

\newcommand{\yr}{\ensuremath{{\rm yr}}}

\newcommand{\new}[1]{{\color{teal}#1}}

\defcitealias{Baldwin1981}{Baldwin, Phillips \& Terlevich}
\defcitealias{Veilleux&Cecil&Bland-Hawthorn2005}{Veilleux, Cecil \& Bland-Hawthorn}
\defcitealias{TumlinsonPeeples&Werk2017}{Tumlinson, Peeples \& Werk}
\newcommand\mycitet[1]{\citetalias{#1}\ (\citeyear{#1})}

\newcommand\mycitept[1]{\citetalias{#1}\ \citeyear{#1}}

\graphicspath{{figs/}}

\title[Drivers of complex emission line profiles in SFGs]{The SAMI Galaxy Survey: $\Sigma_{\rm SFR}$ drives the presence of complex emission line profiles in star-forming galaxies}

\author[H. R. M. Zovaro et al.]{
Henry R. M. Zovaro,$^{1,2}$\thanks{E-mail: henry.zovaro@anu.edu.au}
J. Trevor Mendel,$^{1,2}$
Brent Groves,$^{2,3}$
Lisa J. Kewley,$^{2,4}$
Matthew Colless,$^{1,2}$\newauthor
Andrei Ristea,$^{2,3}$
Luca Cortese,$^{2,3}$
Sree Oh,$^{1,2,5}$
Francesco D'Eugenio,$^{2,6,7}$
Scott M. Croom,$^{2,8}$\newauthor
\'Angel R. L\'opez-S\'anchez,$^{2,9,10}$
Jesse van de Sande,$^{2,8}$
Sarah Brough,$^{2,11}$
Anne M. Medling,$^{1,2,12,13}$\newauthor
Joss Bland-Hawthorn,$^{2,8}$
Julia J. Bryant$^{2,8,14}$
\\
$^{1}$Research School of Astronomy and Astrophysics, The Australian National University, Canberra, ACT 2611, Australia\\
$^{2}$ARC Centre of Excellence for All Sky Astrophysics in 3 Dimensions (ASTRO 3D), Australia\\
$^{3}$International Centre for Radio Astronomy Research, University of Western Australia, 35 Stirling Highway, Crawley WA 6009, Australia\\
$^{4}$Harvard-Smithsonian Center for Astrophysics, 60 Garden Street, Cambridge, MA 02138, USA\\
${^5}$Department of Astronomy, Yonsei University, Seoul 03722, Republic of Korea\\
$^{6}$Kavli Institute for Cosmology, University of Cambridge, Madingley Road, Cambridge, CB3 0HA, United Kingdom\\
$^{7}$Cavendish Laboratory - Astrophysics Group, University of Cambridge, 19 JJ Tohmson Avenue, Cambridge, CB3 0HE, United Kingdom\\
$^{8}$Sydney Institute for Astronomy (SIfA), School of Physics, The University of Sydney, NSW 2006, Australia\\
$^{9}$Australian Astronomical Optics, Macquarie University, 105 Delhi Rd, North Ryde, NSW 2113, Australia\\
$^{10}$Macquarie University Research Centre for Astronomy, Astrophysics \& Astrophotonics, Sydney, NSW 2109, Australia\\
$^{11}$School of Physics, University of New South Wales, NSW 2052, Australia\\
$^{12}$Ritter Astrophysical Research Center, University of Toledo, Toledo, OH 43606, USA\\
$^{13}$Cahill Center for Astronomy \& Astrophysics, California Institute of Technology, 1200 E. California Blvd., Pasadena, CA 91125, USA\\
$^{14}$Australian Astronomical Optics, Astralis-USydney, School of Physics, University of Sydney, NSW 2006, Australia\\
}

\date{Accepted XXX. Received YYY; in original form ZZZ}

\pubyear{2023}

\begin{document}
\label{firstpage}
\pagerange{\pageref{firstpage}--\pageref{lastpage}}
\maketitle

\begin{abstract}

Galactic fountains driven by star formation result in a variety of kinematic structures such as ionised winds and thick gas disks, both of which manifest as complex emission line profiles that can be parametrised by multiple Gaussian components.
We use integral field spectroscopy (IFS) from the SAMI Galaxy Survey to spectrally resolve these features, traced by broad \ha{} components, and distinguish them from the star-forming thin disk, traced by narrow components, in 3068 galaxies in the local Universe. 
Using a matched sample analysis technique, we demonstrate that the presence of complex emission line profiles in star-forming galaxies is most strongly correlated with the global star formation rate (SFR) surface density of the host galaxy measured within 1\re{} (\sfrdens{\re}), even when controlling for both observational biases, including inclination, amplitude-to-noise and angular scale, and sample biases in parameters such as stellar mass and SFR.
Leveraging the spatially resolved nature of the dataset, we determine that the presence of complex emission line profiles within individual spaxels is driven not only by the local $\Sigma_{\rm SFR}$, but by the \sfrdens{\re} of the host galaxy. 
We also parametrise the clumpiness of the SFR within individual galaxies, and find that \sfrdens{\re} is a stronger predictor of the presence of complex emission line profiles than clumpiness.
We conclude that, with a careful treatment of observational effects, it is possible to identify structures traced by complex emission line profiles, including winds and thick ionised gas disks, at the spatial and spectral resolution of SAMI using the Gaussian decomposition technique.

\end{abstract}




\section{Introduction}

The cycling of baryons within and around galaxies plays a critical role in shaping the star formation histories, chemical enrichment, and gas dynamics of present-day galaxies~(e.g., \mycitept{Veilleux&Cecil&Bland-Hawthorn2005}; \mycitept{TumlinsonPeeples&Werk2017}).
In its most basic form, the baryon cycle begins with a galaxy accreting gas from its surroundings---either in the form of a gas-rich merger or directly from the circumgalactic medium---which in turn triggers star formation in the disk, enriching the local interstellar medium with metals.
Ensuing stellar winds and supernovae (SNe) explosions may then expel gas and dust from the disk, creating biconical structures filled with \ha{}-emitting filaments of disk material~\citep{Cooper2008,Cooper2009}. 
Gas entrained by particularly powerful winds can escape the galaxy's gravitational potential, and enrich the intergalactic medium, potentially solving the ``missing baryons'' problem~\citep[e.g.,][]{Shull2012}.
However, in most cases it is expected that the bulk of this material will eventually return to the disk in the form of a ``galactic fountain''~\citep{Shapiro&Field1976,Bregman1980}, redistributing metals throughout the disk and providing fuel from which new generations of stars may form.

Powerful winds launched by vigorous star formation manifest as fast outflows of ionised, neutral, and molecular gas (for a comprehensive overview, see \mycitept{Veilleux&Cecil&Bland-Hawthorn2005}). 
In nearby galaxies such as M82, powerful outflows can be identified by bright, filamentary \ha{} emission in the form of conical structures above and below the disk plane~\citep{Lynds&Sandage1963,Bland&Tully1988,Shopbell&Bland-Hawthorn1998,Sharp&Bland-Hawthorn2010}.
The emission line profiles indicate strong non-circular motions, comprising multiple distinct kinematic components in addition to broad line widths~\citep{McKeith1995,Westmoquette2009a}. 
Studies of large samples of local galaxies have shown that winds are associated with higher stellar masses and star formation rates~\citep[SFRs;][]{Avery2021}, and most strongly with high SFR surface densities~\citep[$\Sigma_{\rm SFR}$;][]{Heckman2002,Roberts-Borsani2020a}. 
More recently, \citet{Law2022} found a strong correlation between the spatially resolved \ha{} velocity dispersion and SFR in local star-forming galaxies, which may be due to increased localised gas turbulence or unresolved broad emission line components at high SFRs.

Many star-forming galaxies do not exhibit winds, but rather thick gas disks that are kinematically distinct from the thin gas disk, which comprises the \hii{} regions.
Present in the majority of disk galaxies~\citep{Haffner2009,Levy2018,Marasco2019} including our own Milky Way~\citep{Reynolds1991}, these thick gas disks---referred to in the literature by a variety of terms, including extraplanar gas or diffuse ionised gas (DIG)---exhibit scale heights of up to a few kpc~\citep{Levy2018,Levy2019,Marasco2019,denBrok2020}, and are therefore substantially thicker than the thin gas disk, but typically have a much lower surface brightness than \hii{} regions.
They also exhibit vertical velocity gradients of between $-10$ and $-30\,\rm km\,s^{-1}\,\kpc^{-1}$\citep{Swaters1997,Levy2018,Levy2019,Marasco2019}, and therefore exhibit a substantial ``lag'' in rotational velocity with respect to the thin disk.
Thick disks have been observed in both \hi{} and \ha{}, with the two phases sharing similar kinematics~\citep[e.g.,][]{Spitzer&Fitzpatrick1993,Reynolds1995,Swaters1997,Howk2003,FraternaliOosterloo&Sancisi2004}, suggesting they represent the same structure. 
Similarly to winds, the presence of thick disks is linked to $\Sigma_{\rm SFR}$~\citep{Rossa&Dettmar2003a,Rueff2013,Levy2018}.

\citet{Fraternali&Binney2006} proposed that thick ionised gas disks may form via gas return from galactic fountains. 
Indeed, the kinematics of the thick disk are incompatible with simple hydrostatic equilibrium~\citep{Barnabe2006,Marinacci2010a}, and require continuous injection of turbulence to remain stable. 
Furthermore, recent hydrodynamical simulations suggest that these thick disks may form due to fountains powered by star formation evenly distributed within the disk, whereas centrally concentrated star formation is more likely to produce powerful outflows~\citep{Vijayan2018}.
Accretion has also been proposed as a mechanism for the formation of thick disks, although this probably only contributes in a small fraction of galaxies at low redshift~\citep{Marasco2019}.
Understanding the origins of these thick disks, and their relation to winds, therefore provides insight into how gas is cycled in and out of star-forming galaxies.

In high-resolution spectroscopy, winds and thick disks can be identified by the presence of complex emission line profiles. Whilst winds will regularly produce clear ``line splitting'' and broad emission line components~\citep[e.g.,][]{Westmoquette2009a}, thick disks present as subtle wings tracing gas at lower rotational speeds than the thin disk~\citep[e.g.,][]{Marasco2019}. 
Indeed, \citet{Belfiore2022} proposed that thin and thick disk \ha{} emission can be separated via fitting multiple Gaussian emission line components; this was demonstrated earlier by \citet{denBrok2020}, who distinguished thick disk emission in a small sample of local galaxies by fitting multiple Gaussian line profiles to the \ha{} line profiles in low-spectral resolution MUSE observations. 

The Sydney-Australian Astronomical Observatory (AAO) Multi-object Integral field spectrograph (SAMI) Galaxy Survey~\citep{Bryant2015,Croom2021} provides optical integral field spectroscopy (IFS) for 3068 local galaxies. Crucially, the spectral resolution of the red arm of AAOmega~\citep[$\sigma_{\rm inst} = 29.6\,\rm km \, s^{-1}$;][]{vandeSande2017b,Scott2018} is sufficient to resolve individual Gaussian components in the emission lines from ionised gas~\citep[e.g.,][]{Ho2014}, thereby enabling detailed studies of spectrally resolved winds and thick disks in a large sample of galaxies.

In this work, we use the SAMI Galaxy Survey to determine the underlying galaxy properties associated with complex emission profiles in star-forming galaxies, and therefore the processes that lead to the formation of galactic fountains, winds and thick ionised gas disks. 
In Section~\ref{sec: the SAMI galaxy survey} we introduce the sample, and in Section~\ref{sec: Defining 1- and multi-component galaxies} we detail how we identify galaxies with and without complex emission line profiles. In Section~\ref{sec: Matched sample analysis} we conduct a matched sample analysis to determine the underlying galaxy properties most strongly associated with complex emission line profiles, and in Section~\ref{sec: local vs. global effects} we investigate the roles of star formation on both local and global scales. Finally, we discuss our results plus the influence of observational biases in Section~\ref{sec: Discussion}.
For the remainder of this work we assume a flat $\Lambda$CDM cosmology with $H_0 = 70\,\rm km\,s^{-1}\,Mpc^{-1}$, $\Omega_{\rm M} = 0.3$ and $\Omega_{\Lambda} = 0.7$.

%
%

\section{The SAMI Galaxy Survey}\label{sec: the SAMI galaxy survey}

The SAMI Galaxy Survey~\citep{Croom2012,Bryant2015} is an IFS survey of 3068 local ($z < 0.095$) galaxies spanning a stellar mass range $10^{7} - 10^{12} \msol$ in field, group and cluster environments.
Each target was observed with the Sydney Australian Astronomical Observatory Multi-Object Integral Field Spectrograph~\citep[SAMI;][]{Croom2012} on the 3.9\,m Anglo-Australian Telescope at Siding Spring Observatory.
SAMI is a multi-object fused-fibre-bundle instrument that feeds the AAOmega spectrograph~\citep{Sharp2006}, which has separate blue (3700--5700\,\AA, $R = 1812$) and red (6300--7400\,\AA; $R = 4263$) arms. 
Each fibre bundle, or \textit{hexabundle}~\citep{Bland-Hawthorn2011,Bryant2014}, comprises 61 individual 1.6'' diameter optical fibres with a high filling factor of 75\,per\,cent and a total footprint of 14.7''.
The observations for each target were dithered on-source to account for gaps between fibres in the hexabundle. 
The final reduced data cubes have 0.5'' $\times$ 0.5'' \textit{spaxels} (spatial pixels) with an average effective seeing of 2''.

We used the data products from the third SAMI data release~\citep[DR3;][]{Croom2021}\footnote{Available at \url{https://docs.datacentral.org.au/sami/}.}, which includes the original data cubes, plus two-dimensional maps of emission line fluxes, gas and stellar kinematics, and other quantities.
DR3 also includes tables\footnote{Available at \url{https://datacentral.org.au/services/schema/\#sami}.} containing measurements of additional physical quantities and data quality flags for each galaxy.
Spectroscopic and flow-corrected redshifts ($z$) and stellar masses ($M_*$) were extracted from the \texttt{InputCatGAMADR3}~\citep{Bryant2015}, \texttt{InputCatClustersDR3}~\citep{Owers2017} and \texttt{InputCatFiller} tables. 
Effective radii (\re{}) and axis ratios ($b/a$) were derived from $r$-band SDSS, VST and GAMA photometry were computed using the multi-Gaussian expansion~\citep[MGE;][]{Emsellem1994,Cappellari2002} fits described in \citet{DEugenio2021}. 
Inclinations ($i$) were derived from these axis ratios using eqn. 1 from \citet{Poetrodjojo2021} adopting an intrinsic disk thickness of $q_0 = 0.2$.

To obtain spatially resolved measurements for each galaxy, we used the unbinned (``default'') data products, which include emission line profile fits carried out using \textsc{lzifu}~\citep{Ho2016b}, which first fits and then subtracts the stellar continuum to more accurately fit the emission lines.
In this work we use the ``recommended'' component fits, in which \lzcomp{}~\citep{Hampton2017a,Hampton2017b}, an artificial neutral network (ANN), was used to evaluate the best-fitting number of components in each spaxel (i.e., one, two or three components).
The components are ordered by velocity dispersion, such that components 1, 2 and 3 refer to the narrow, broad and extra-broad components respectively. 
For each galaxy, DR3 includes maps of the \textit{total} emission line fluxes in each spaxel---i.e., summed over all Gaussian emission line components---for \forb{O}{ii}$\uplambda \uplambda 3726,9$, \hb{}$\uplambda 4861$, \forb{O}{iii}$\uplambda\uplambda 4959,5007$, \forb{O}{i}$\uplambda 6300$, \forb{N}{ii}$\uplambda\uplambda 6548,83$, \ha{}$\uplambda 6563$ and \forb{S}{ii}$\uplambda\uplambda 6716,31$; fluxes for individual components are only provided for \ha{} due to S/N limitations and the lower spectral resolution of the blue arm of AAOmega. The line-of-sight (LoS) ionised gas velocity (\vgas{}) and velocity dispersion (\sigmagas{}, corrected for instrumental resolution by \textsc{lzifu}) are provided for each individual component, however.

We also used the SFR and $\Sigma_{\rm SFR}$ maps produced by \citet{Medling2018}, which were derived using the reddening-corrected \ha{} fluxes from the recommended-component fits assuming a \citet{Chabrier2003} initial mass function (IMF) and the SFR calibration of \citet{Kennicutt1994}.
We note that these SFR measurements are based on the total \ha{} emission line flux, which may lead to over-estimates of the SFR in cases where broad emission line components are associated with emission from other sources such as shocks or DIG. 
However, \citet{Belfiore2022} argue that in nearby star-forming galaxies, DIG is predominantly ionised by leaky \hii{} regions, and should therefore be included in SFR estimates; they also concluded that including DIG from evolved stellar populations in these estimates will not lead to a significant over-estimate of the SFR in kpc-resolution data, such as SAMI. 
None the less, we repeated our analysis using SFR measurements from only the narrow \ha{} component in each spaxel, and found that it made no meaningful difference to our conclusions.
Global SFRs and $\Sigma_{\rm SFR}$ were measured for each galaxy by summing the SFRs in these maps within nuclear apertures of diameter 3\,kpc and $2\re$.

Emission line fluxes were corrected for extinction using the \citet{Fitzpatrick&Massa2007} reddening curve with $R_V = 3.1$ and assuming an intrinsic Balmer decrement of $\ha{}/\hb{} = 2.86$ which corresponds to a standard nebular electron temperature and density of $10^{4}\,\rm K$ and $100\,\rm cm^{-3}$ respectively~\citep{Dopita&Sutherland2003}. To ensure reliable extinction correction, this step was only applied to spaxels in which the total flux S/N of both \ha{} and \hb{} exceeded 5. \ha{} equivalent widths (\ewha{}) were computed as the non-extinction-corrected \ha{} flux divided by the mean continuum level in the rest-frame wavelength range 6500--6540\,\AA.

\subsection{Data quality and S/N cuts}\label{subsec: Data quality and S/N cuts}

Using the DR3 \texttt{CubeObs} table as described in \citet{Croom2021}, we identified and removed galaxies with bad sky subtraction residuals (as indicated by \texttt{WARNSKYR} and \texttt{WARNSKYB}), those with flux calibration issues (\texttt{WARNFCAL} and \texttt{WARNFCBR}), and those containing multiple objects in the SAMI field-of-view (FoV) (\texttt{WARNMULT}), leaving 2997 galaxies. 

To ensure high-quality kinematics and \ha{} fluxes of individual components within individual spaxels, we adopted a minimum flux S/N (defined as the flux divided by the formal uncertainty provided by \lzifu) of 5 in \ha{} in each component, plus a component amplitude-to-noise (A/N) of at least 3, here defined as the Gaussian amplitude of the component divided by the RMS noise in the rest-frame wavelength range 6500\,Å--6540\,Å.
We also required a minimum gas velocity dispersion S/N of 3 in each component as defined by \citet{Federrath2017}, which accounts for the uncertainty in measuring the width of individual emission line components that are close to the instrumental resolution. 
For total emission line fluxes in each spaxel (that is, summed over all emission line components), we required a minimum S/N of 5. 
Example fits from spaxels meeting these criteria are provided in Appendix~\ref{app: Example LZIFU fits}.

\subsection{Spectral classification}\label{subsec: Spectral classification}

Spaxels were spectrally classified using the standard optical diagnostic diagrams of \mycitet{Baldwin1981} and \citet{Veilleux&Osterbrock1987}, which plot the \forb{O}{iii}$\uplambda 5007$/\hb{} ratio (O3) against the \forb{N}{ii}$\uplambda 6583$/\ha{} (N2), \forb{S}{ii}$\uplambda\uplambda 6717,31$/\ha{} (S2), and \forb{O}{i}$\uplambda 6300$/\ha{} (O1) ratios.

We used only the O3 vs. N2 and O3 vs. S2 diagrams due to the relative weakness of the \forb{O}{i} emission line. 
To ensure reliable classification, only spaxels with a flux S/N of at least 5 in all of \hb{}, \ha{}, \forb{O}{iii}$\uplambda 5007$, \forb{N}{ii}$\uplambda 6583$, and \forb{S}{ii}$\uplambda\uplambda 6716,31${} lines were classified.
Due to the lower spectral resolution of the blue arm ($R = 1812$) compared to the red arm ($R = 4236$), blue lines such as \forb{O}{iii} and \hb{} may not have reliable flux measurements for individual components. As a result, DR3 only provides total fluxes for lines other than \ha{}, rather than those of individual components. Spectral classification was therefore performed using the total emission line fluxes in each spaxel. 
We note that, in cases where multiple emission line components are present that arise from different excitation mechanisms (e.g., a star-forming component plus a shocked component), the classification is naturally weighted towards the component with the highest luminosity.

Each spaxel was assigned one of the following spectral classification:
\begin{itemize}
	\item \textbf{Star-forming (SF)}: lies below the \citet{Kauffmann2003} line in the N2 diagram, and below the \citet{Kewley2001} ``extreme starburst'' line in the S2 diagram.
	\item \textbf{Composite}: lies above the \citet{Kauffmann2003} line in the N2 diagram but below the the \citet{Kewley2001} ``extreme starburst'' line in both the N2 and S2 diagrams.
	\item \textbf{LINER}: lies above the \citet{Kewley2001} ``extreme starburst'' line in both the N2 and S2 diagrams but below the \citet{Kewley2006} LINER/Seyfert line in the S2 diagram.
	\item \textbf{Seyfert}: lies above the \citet{Kewley2001} ``extreme starburst'' line in both the N2 and S2 diagrams but above the \citet{Kewley2006} LINER/Seyfert line in the S2 diagram.
	\item \textbf{Ambiguous}: inconsistent classifications between the O3 vs. N2 and O3 vs. S2 diagrams; e.g., composite-like in the N2 diagram, but LINER-like in the S2 diagram.
	\item \textbf{Not classified}: low S/N in or missing at least one of \ha{}, \hb{}, \forb{N}{ii}, \forb{S}{ii} or \forb{O}{iii} fluxes.
\end{itemize}

\section{Defining 1- and multi-component galaxies}\label{sec: Defining 1- and multi-component galaxies}

\begin{figure*}
	\centering
	\includegraphics[width=\linewidth]{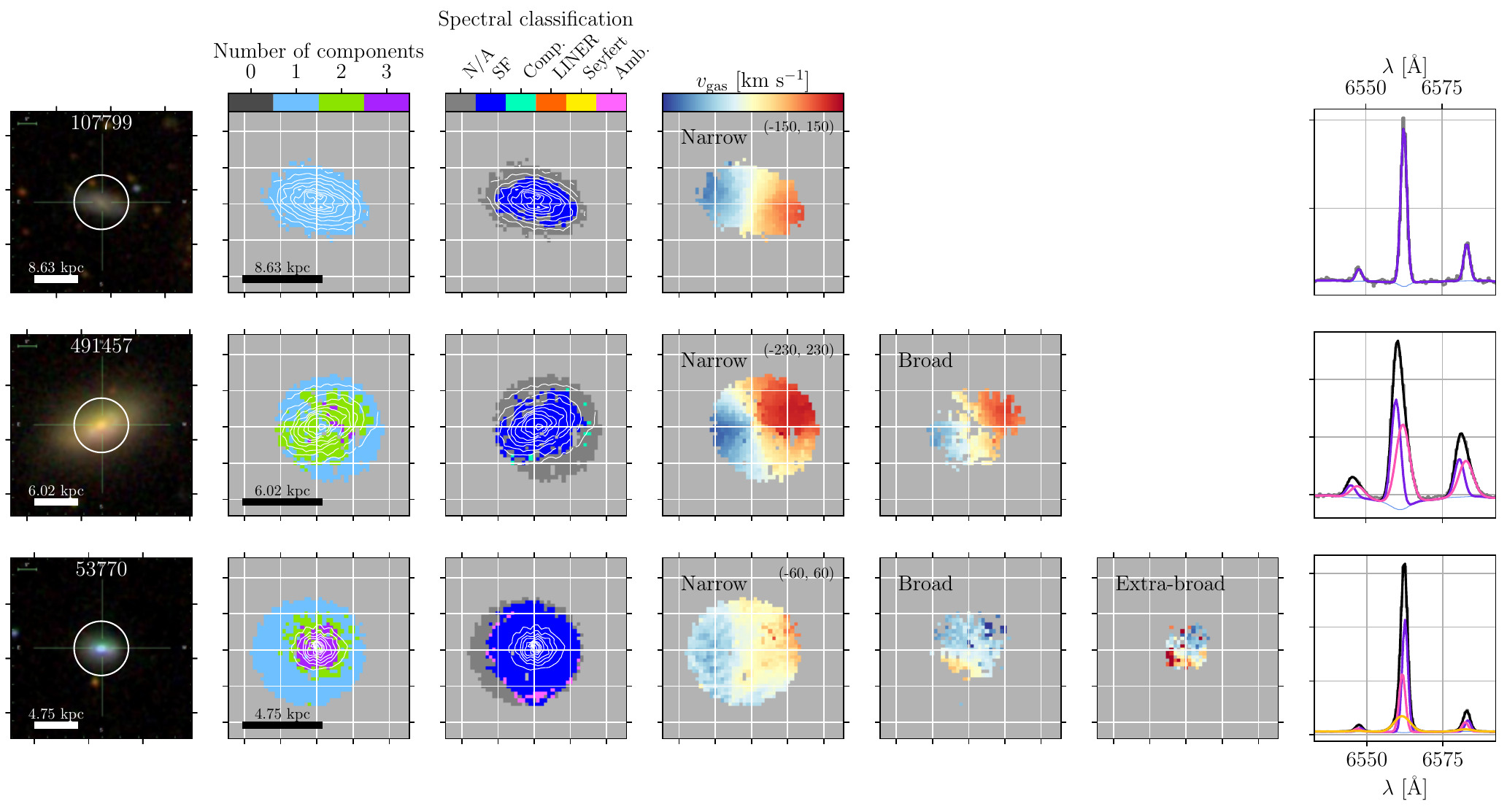}
	\caption{
		Example 1-component (top row) and multi-component (bottom two rows) galaxies. The leftmost column shows the SDSS image, where the white circle represents the SAMI FoV. The middle five columns show maps of the number of components, spectral classification, and LoS velocities of the ionised gas in each spaxel, where the white contours represent the logarithmically-scaled stellar continuum, and the black bar shows the angular scale. The minimum and maximum velocity scaling values for each galaxy are shown in the upper-right corner.
		The rightmost column shows the rest-frame \ha{} and \forb{N}{ii}$\uplambda\uplambda 6548,83$ emission line profiles in grey, overlaid with the \lzifu{} fits, extracted from an individual spaxel in the nucleus of each galaxy. 
		The blue line represents the stellar continuum fit; the purple, pink and orange lines represent the first (narrow), second (broad) and third (extra-broad) Gaussian components respectively, and the black line represents the combined fit.
	}
	\label{fig: example galaxies}
\end{figure*}   

Our aim is to determine the fundamental differences between star-forming galaxies exhibiting large numbers of spaxels with complex emission line profiles---that is, containing multiple components---and those without, which we refer to as ``multi-component galaxies'' and ``1-component galaxies'' respectively. Here, we describe our method for classifying SAMI galaxies into these two categories.

We begin by defining a parent sample. For a galaxy to be included, it must contain at least 50 spaxels that can be spectrally classified after making the data quality and S/N cuts detailed in Section~\ref{subsec: Data quality and S/N cuts}; this requirement removes galaxies that are poorly resolved and those with overall low emission line S/N and therefore less reliable multi-component fits. 
In this work, we are primarily interested in winds and outflows associated with star formation. Because active galactic nuclei (AGN) can also contribute to these phenomena, we remove possible AGN hosts from our sample by requiring at least 80\,per\,cent of spectrally-classified spaxels to be star-forming, and that all but one spaxel within 2'' of the nucleus must be star-forming, unclassified, or have $\ewha < 3\,\AA$, where this last criterion allows for spaxels where the dominant photoionisation mechanism is evolved stars~\citep{Singh2013,Belfiore2016,Byler2019}. 
We also remove galaxies containing any spaxels classified as Seyfert. 
Our final parent sample of star-forming galaxies contains \textbf{655} objects.

From this parent sample, we select ``1-component'' and ``multi-component'' galaxies according to the following criteria: 1-component galaxies are those in which over 99\,per\,cent of all spaxels with emission line components that pass our S/N and data quality cuts contain only 1 component, and multi-component galaxies are those in which at least 15\,per\,cent of all such spaxels contain 2 or more components. Objects in which 1--15\,per\,cent of spaxels contain multiple components are not classified as either 1- or multi-component galaxies.

Figure~\ref{fig: example galaxies} shows examples of 1- and multi-component galaxies. SDSS images are shown, alongside maps displaying the number of emission line components fitted in each spaxel, spectral classifications, and the LoS velocity of the narrow and broad components. 
Note that we do not base our classification on the number of 3-component spaxels due to their rarity: of all spaxels containing emission lines exceeding our S/N requirements, approximately 6.8\,per\,cent contain 2 components whereas only 0.7\,per\,cent contain 3 components. 
That said, our multi-component galaxy sample contains objects with appreciable numbers of spaxels containing 3 components, e.g., galaxy 53770 as shown in Fig.~\ref{fig: example galaxies}. 

Although our choice of 15\,per\,cent is somewhat arbitrary, we find that shifting this threshold does not meaningfully change our results; 15\,per\,cent strikes a good balance between having a sufficiently large multi-component galaxy sample to conduct statistical analysis, and having a large enough number of multi-component spaxels within each galaxy to ensure our multi-component sample is not contaminated by spaxels with unreliable 2 or 3-component fits. 
Our criteria result in \textbf{266} 1-component galaxies and \textbf{133} multi-component galaxies. 

Note that our selection criteria for multi-component galaxies do not explicitly take into account the origin of the broad-line emission, and as such includes systems in which the broad component is due to beam smearing, winds/outflows, thick disks and other processes such as mergers and stripping. For example, as shown in Fig.~\ref{fig: example galaxies}, the broad component in galaxy 491457 exhibits smooth rotation similar to that of the narrow component, whereas that in galaxy 53770 is clearly blueshifted, and likely represents a wind.

A drawback of spatially resolved data in comparison with integrated spectra is the lower S/N. 
We therefore repeated our 1- and multi-component classification process using velocity-subtracted 1\re{} aperture spectra to check whether there are any galaxies with low surface-brightness broad line emission that is not detectable in the spatially resolved data. As detailed in Appendix~\ref{app: Unresolved selection criteria}, there were no galaxies classified as 1-component using the resolved data that were classified as multi-component in the aperture spectra. In fact, 20 galaxies classified as multi-component using the resolved data were classified as 1-component using the aperture spectra, possibly due to beam smearing effects ``washing out'' faint broad components in individual spaxels. We therefore rely only upon the resolved classifications for the remainder of this paper.

\subsection{General properties of 1- and multi-component galaxies}\label{subsec: General properties of 1- and multi-component galaxies}
\begin{figure*}
	\centering
	\begin{subfigure}[b]{\linewidth}
		\includegraphics[width=1\linewidth]{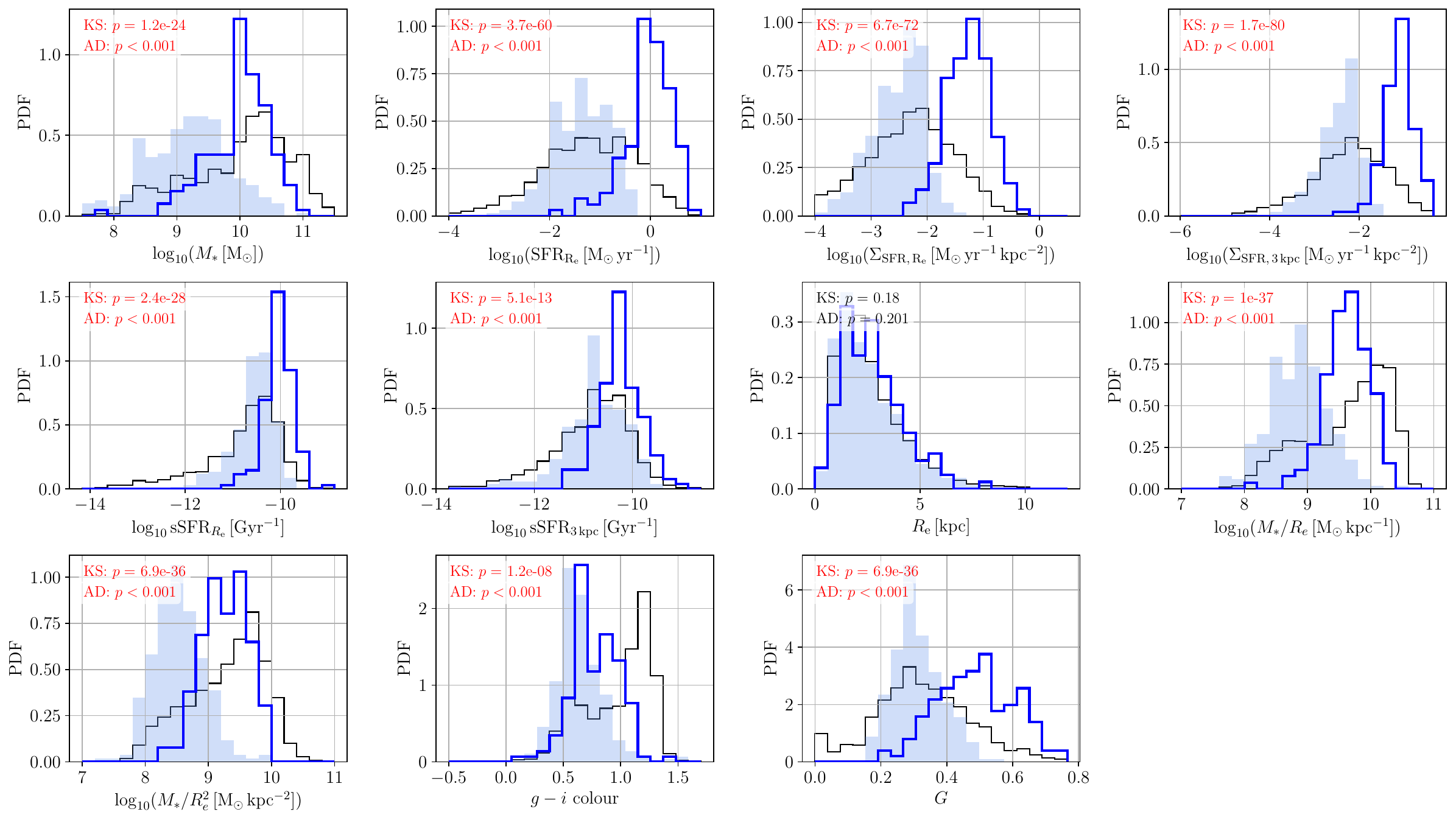}
		\caption{}
		\label{fig: sample histograms (physical)}
	\end{subfigure}
	\begin{subfigure}[b]{\linewidth}
		\includegraphics[width=1\linewidth]{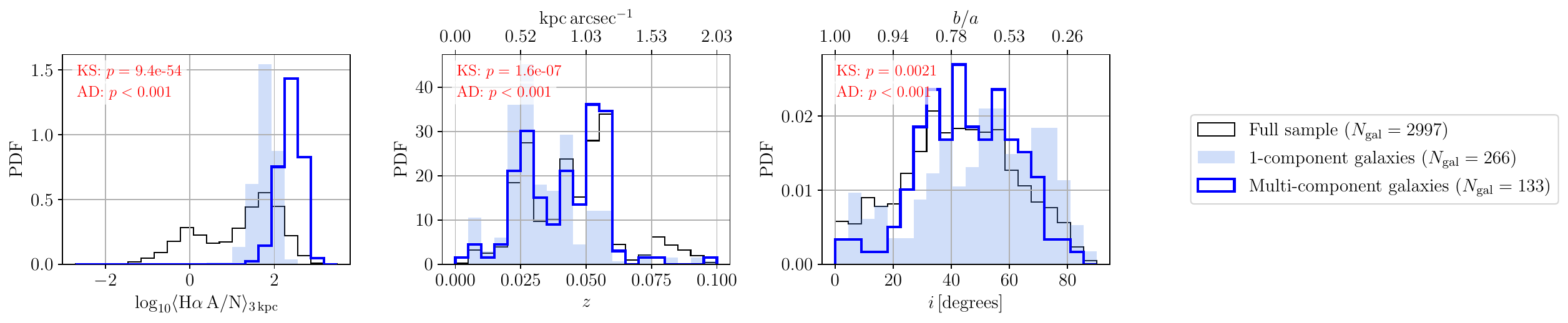}
		\caption{}
		\label{fig: sample histograms (observational)}
	\end{subfigure}
	\caption{
		Probability distribution functions (PDFs) in various \textit{physical} galaxy properties (a) and \textit{observational} quantities (b) shown for our different sub-samples. In each panel we show the distribution in the full SAMI sample of 2997 galaxies (thin black line), the 1-component galaxy sample (shaded blue histogram), and multi-component galaxy sample (thick blue line). 
		In (b), the top axis ticks of the bottom right-most panel show the axis ratios $b/a$ corresponding to the measured inclinations assuming an intrinsic disk thickness $q_0 = 0.2$~\citep{Poetrodjojo2021}.
		The text in the upper left corner shows the $p$-values of the KS and AD tests applied to the 1- and multi-component samples. In all quantities except for \re, the very low $p$-values indicate that the samples are highly unlikely to be drawn from the same parent sample, suggesting the two samples have fundamentally different distributions in all quantities shown here.}
	\label{fig: sample histograms}
\end{figure*}

Figures~\ref{fig: sample histograms (physical)} and \ref{fig: sample histograms (observational)} show the distributions of various physical and observational galaxy properties respectively in the 1- and multi-component samples. 
The 2-sample Kolmogorov-Smirnoff (KS) and Anderson-Darling (AD) tests, both of which estimate the likelihood that both samples are drawn from the same parent distribution, were applied to the 1- and multi-component galaxy distributions in each quantity. In all quantities except for \re{}, the $p$-values are far below the significance threshold of $p = 0.05$, suggesting fundamental differences between the 1- and multi-component galaxies.
Both tests are employed in order to more robustly judge whether distributions are significantly different or not; in cases where one test has $p > 0.05$ and the other has $p < 0.05$, we judge the distributions as being substantially different only at a tentative level.

As shown in Fig.~\ref{fig: sample histograms (physical)}, the multi-component galaxies are more massive, and have significantly higher mean SFRs and $\Sigma_{\rm SFR}$ measured within 1\re{} (\sfr{\re} and \sfrdens{\re}, respectively) as well as higher specific SFRs (sSFRs) than 1-component galaxies. 
In particular, there is essentially no overlap in the distributions in $\Sigma_{\rm SFR}$ measured within 3\,kpc (\sfrdens{{\rm 3\,kpc}}) of the 1- and multi-component systems; however, \sfrdens{\rm 3\,kpc} may not reflect the true $\Sigma_{\rm SFR}$ within the galaxy in systems with $\re < 3\,\kpc$, which represents a significant fraction of our galaxies. For this reason, we primarily rely on \re{}-based measurements for the remainder of this work.
The multi-component galaxies also have higher values of $\log_{10}(\mstar/\re)$ and $\log_{10}(\mstar/\re^2)$, which are proxies for gravitational potential and stellar surface density respectively, and slightly redder $g-i$ colours than 1-component galaxies. They also have clumpier star formation distributions on average, as evidenced by their higher Gini coefficients~\citep[$G$,][]{Gini1936}, which are computed using 
\begin{equation}
	G = \frac{1}{2 n^2 \bar{s}}\sum_{i=1}^{n}\sum_{j=1}^{n}|s_i - s_j|
	\label{eq: Gini coefficient}
\end{equation}
where $s_i$ is the $\Sigma_{\rm SFR}$ (or equivalently SFR) in spaxel $i$, $n$ is the total number of spaxels where the $\Sigma_{\rm SFR}$ can be measured and $\bar{s}$ is the average $\Sigma_{\rm SFR}$ in the galaxy. $G = 0$ corresponds to perfect equality, i.e., every spaxel has the same $\Sigma_{\rm SFR}$, whereas $G = 1$ corresponds to maximal inequality, i.e., all star formation is concentrated in a single spaxel. 
Gini coefficients are discussed further in Section~\ref{sec: local vs. global effects}.

Figure~\ref{fig: sample histograms (observational)} shows the distribution in various observational quantities in the 1- and multi-component samples. 

In the leftmost panel we show $\log_{10} \langle \ha \, {\rm A/N} \rangle_{\rm 3kpc}$, the mean \ha{} A/N measured within a 3\,kpc aperture. Here, and for the remainder of this paper, \ha{} A/N is defined within each spaxel as the peak amplitude of the \ha{} line profile (directly measured from the spectrum, and therefore not corrected for stellar absorption) relative to the mean continuum level in the rest-frame wavelength range 6500\,Å--6540\,Å divided by the RMS noise measured in the same range. We opt to use A/N rather than flux S/N as the latter is based on the fluxes and formal uncertainties provided by \lzifu{}, and may not representative of the true S/N in cases of bad fits.
The multi-component galaxies exhibit higher mean \ha{} A/N than the 1-component galaxies.

They also tend to be more distant, although this could result from SAMI's stepped sample selection function, in which all essentially all galaxies with $\log_{10} \mstar < 9$ are located at $z \lesssim 0.03$~\citep[see fig.\,3 of][]{Croom2021}.
Compared to the 1-component galaxies, multi-component galaxies are slightly over-represented at intermediate inclinations and under-represented at high inclinations, which is contrary to expectations if the presence of multiple components is a consequence of beam smearing alone. Potential explanations for this over-representation are discussed in Section~\ref{subsec: Observational biases}.

\begin{figure}
	\centering
	\includegraphics[width=1\linewidth]{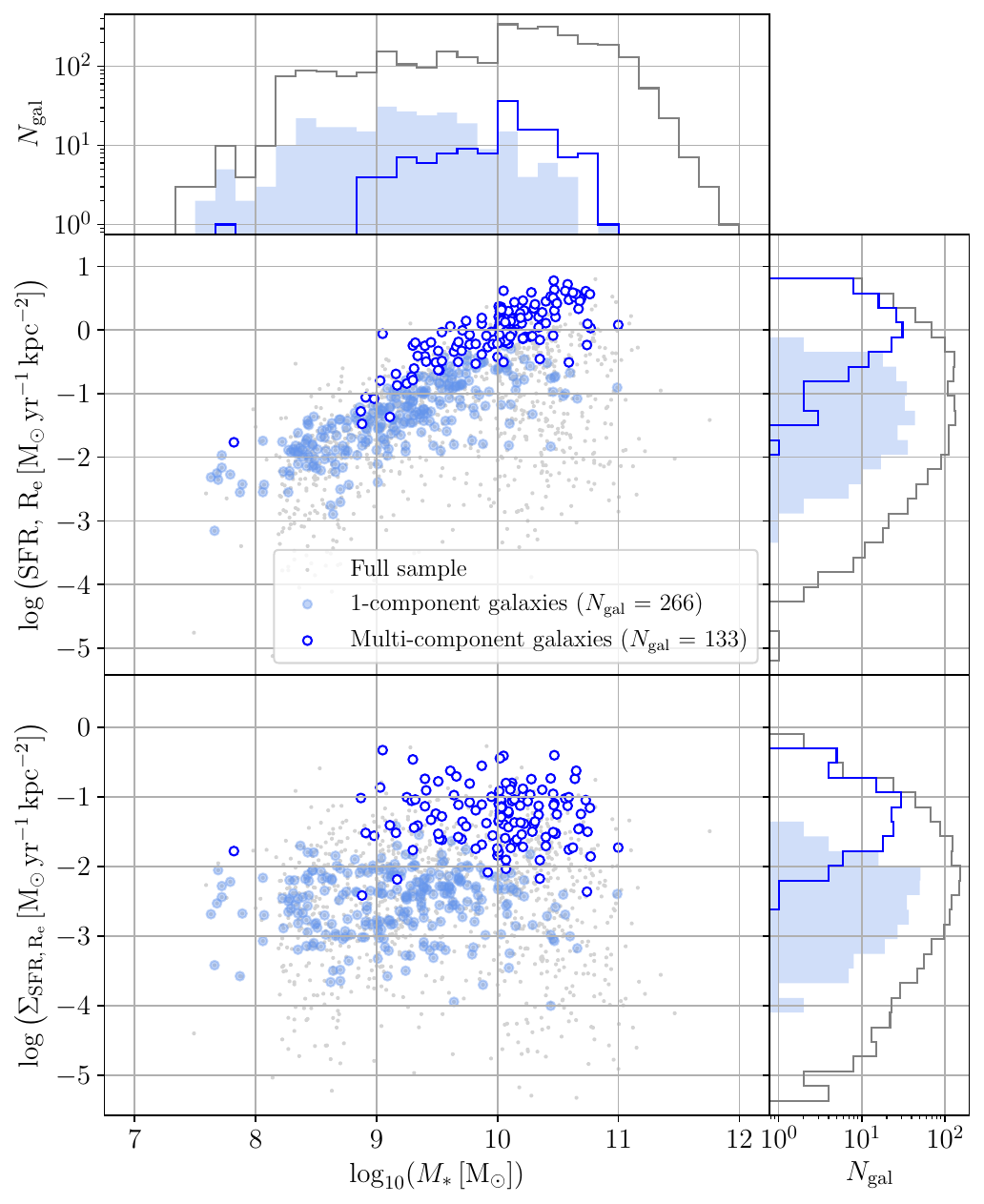}
	\caption{
		Star-forming main sequence (top) and $\Sigma_{\rm SFR}$ main sequence (bottom) for the full SAMI sample (grey points), 1-component galaxies (blue filled circles) and multi-component galaxies (blue open circles). Distributions in \mstar{}, \sfr{\re} and \sfrdens{\re} for each subsample are shown in the top and right-hand panels.
	}
	\label{fig: SFMS/SFR surface density MS}
\end{figure}

The locations of the 1- and multi-component galaxies on the star-forming main sequence (SFMS) are shown in the upper panel of Fig.~\ref{fig: SFMS/SFR surface density MS}. At stellar masses below approximately $10^{10}\,\msol$, multi-component galaxies sit along the upper envelope of the SFMS, having elevated SFRs for their mass, whilst at higher masses they exhibit a wider range of SFRs. 
The bottom panel, which shows \sfrdens{\re} as a function of stellar mass, shows that the multi-component galaxies typically have $\Sigma_{\rm SFR} \gtrsim 10^{-2} \, \msol \, \yr^{-1} \, \kpc^{-2}$.

The \re{} measurements were derived using $r$-band photometry~\citep[for details see][]{DEugenio2021} which includes the \ha{} emission line for galaxies at $z \lesssim 0.066$, representing the majority of galaxies in our sample (see Fig.~\ref{fig: sample histograms}b). In galaxies with high SFRs, \re{} may be biased towards smaller values by bright \ha{} emission in the nuclear regions, thereby no longer representing the stellar extent of the galaxy and in turn elevating \sfrdens{\re}.
A 10\,per\,cent \ha{} contribution to the $r$-band flux corresponds to an emission line \ewha{} of approximately 150\,\AA; within 1\re{}, the average \ewha{} exceeds this amount in only 3 and 2 of our 1- and multi-component galaxies respectively, indicating that only a very small fraction of our sample is likely to have biased \re{} measurements. We therefore conclude that the \re{} measurements are not significantly biased by this effect and use \re{}-aperture measurements for the remainder of this work. Plots made using the 3\,kpc circular (``round'') aperture measurements are shown in Appendix~\ref{app: additional plots and tables} for comparison. 


\section{Matched sample analysis}\label{sec: Matched sample analysis}

As shown by Fig.~\ref{fig: sample histograms}, the 1- and multi-component galaxies have statistically significantly different distributions in a number of physical properties, including stellar mass and SFR; they also differ in observational properties such as inclination and S/N. 
Identifying the processes that cause complex emission line profiles therefore requires careful treatment due to underlying correlations between various parameters. 
For example, the multi-component galaxies have higher stellar masses and higher SFRs; however, because stellar mass is itself correlated with SFR (Fig.~\ref{fig: SFMS/SFR surface density MS}), this alone does not indicate whether the presence of complex emission line profiles is driven by stellar mass or SFR. 

Observational biases must also be controlled for. For instance, the multi-component galaxies have higher \ha{} A/N; a scenario in which \textit{all} galaxies would exhibit multiple emission line components given sufficient A/N could therefore be consistent with these observations. However, the multi-component galaxies also have higher SFRs, meaning that at a fixed redshift they will have brighter \ha{} emission, and therefore higher A/N. The offset in A/N could therefore also occur if SFR is the fundamental driver of multiple emission line components.

To control for the effects of observational and physical parameters on the presence of multiple emission line components, we perform a matched sample analysis~\citep[e.g.,][]{Kaasinen2017} using an adaptive threshold matching technique as follows:

\begin{enumerate}
	\item For each multi-component galaxy, we identify all unmatched 1-component galaxies within some matching threshold $\Delta$ of the quantity $Q$ being matched. \label{matching step: 1}
	\item If this is successful, we select the 1-component galaxy from this subset that is closest to the multi-component galaxy in $Q$, and subsequently remove both galaxies from the lists of unmatched galaxies. 
	\item Otherwise, if there are no 1-component galaxies within $\pm \Delta$ of $Q$, the multi-component galaxy remains unmatched and we repeat step \ref{matching step: 1} for the next unmatched multi-component galaxy.
	\item Once we have attempted to match all multi-component galaxies, we terminate the process if at least 80\,per\,cent of all multi-component galaxies have been matched. Otherwise, we increment the matching threshold by $+\Delta$ and repeat the matching process (step \ref{matching step: 1}) on all of the unmatched multi-component galaxies.
	\item If necessary, steps \ref{matching step: 1}--\ref{matching step: 5} are repeated once more with a match threshold of $3\Delta$, after which the process is terminated. \label{matching step: 5}
\end{enumerate}

$\Delta$ was set to the typical uncertainty for quantities in which measurement uncertainties were available; $\Delta$ values for other parameters were chosen to provide sufficiently large matched samples to enable fair comparison whilst ensuring reasonable similarity in the matched quantity between 1- and multi-component galaxies. The values of $\Delta$ used are given in Table~\ref{tab: matched sample thresholds}.

\begin{table}
	\caption{Match thresholds $\Delta$ used for the matched sample analysis (Section~\ref{sec: Matched sample analysis}).}
	\centering
	\begin{tabular}{cc}
		\hline 
		\textbf{Quantity} & \textbf{$\Delta$} \\ 
		\hline 
		$\log_{10} (M_*\,[\rm M_\odot])$ & 0.1 \\
		$\log_{10} (\sfr{\re}\, [\rm M_\odot\,\yr^{-1}])$ & 0.0085\\
		$\log (\sfrdens{\re}\, [\rm M_\odot\,\yr^{-1}\,\kpc^{-2}])$ & 0.0085 \\
		$G({\rm SFR})$ & 0.025 \\
		\re{} [kpc] & 0.2 \\
		$\log_{10}(M_*/R_e\,[\rm M_\odot\,kpc^{-1}])$ & 0.1 \\
		$\log_{10}(M_*/R_e^2\,[\rm M_\odot\,kpc^{-2}])$ & 0.1 \\
		$i$ [degrees] & 2 \\
		kpc arcsec$^{-1}$ & 0.1 \\
		$\log_{10} \langle {\rm H\upalpha} \, \rm A/N \rangle_{\rm 3\,kpc} $ & 0.05 \\
		\hline
	\end{tabular}
	\label{tab: matched sample thresholds}
\end{table}


Validity of the final matched 1- and multi-component galaxy samples was determined using the 2-sample KS and AD tests as in Section~\ref{subsec: General properties of 1- and multi-component galaxies}; when matching, we required $p > 0.05$ in both of these tests to ensure that the matched samples were valid.
To ensure stability of our results, we repeated steps \ref{matching step: 1} to \ref{matching step: 5} 1000 times for each matched quantity, each time randomly sorting the list of multi-component galaxies, and inspecting the distribution of $p$-values. In all matched quantities the $p$-values exceeded 0.05 in both KS and AD tests over 99\,per\,cent of the time, indicating stable results.
		
\subsection{Matched sample analysis results} 

Figure~\ref{fig: matched sample analysis - violin plots} shows violin plots representing kernel density estimates (KDEs) of the distributions in various parameters for the 1- and multi-component galaxy samples matched in \mstar\, \sfr{\re}, \sfrdens{
\re} and $G$. 
To control for observational biases, we also show samples simultaneously matched with these physical quantities and in succession with $\log_{10} \langle {\rm H\upalpha} \, \rm A/N \rangle_{\rm 3\,kpc} $, angular scale and inclination.

To control for observational biases, we also show samples matched with these physical quantities simultaneously in $\log_{10} \langle {\rm H\upalpha} \, \rm A/N \rangle_{\rm 3\,kpc} $, angular scale and inclination. 
The results of KS and AD 2-sample tests are indicated for each violin plot; text in red represent $p < 0.05$, indicating a statistically significant likelihood that the samples are not drawn from the same underlying distribution. 
Results for matched samples in other combinations of parameters are summarised in Fig.~\ref{fig: matched sample analysis - punnet square}.

We first match in \mstar\ (first four columns of Fig.~\ref{fig: matched sample analysis - violin plots}). When controlling for \mstar{} (column 1), the multi-component galaxies have smaller \re{} and significantly higher \sfr{\re}; in particular, they exhibit mean \sfrdens{\re} and \sfrdens{{\rm 3\,kpc}} nearly 1.0 dex higher than those of the matched 1-component galaxies.
They also have significantly higher values of $G$, indicating clumpier and/or more centrally-concentrated SFR distributions, much higher average \ha{} A/N and have distinct distributions in angular scale and inclination.
However, when controlling for \ha{} A/N, angular scale and inclination by matching in these quantities simultaneously with \mstar\ (columns 2--4), the multi-component galaxies still exhibit significantly higher \sfr{\re}, $\Sigma_{\rm SFR}$ and $G$, and tend to be more compact.

We now consider \sfr{\re} (columns 5--8 of Fig.~\ref{fig: matched sample analysis - violin plots}). 
When matched in \sfr{\re} alone (column 5), the multi-component galaxies have a significantly different distribution in \mstar, being less massive than the 1-component galaxies; however, when controlling for \ha\ A/N, angular scale and inclination this difference disappears. Rather, when accounting for these observational effects, the multi-component galaxies are differentiated by being more compact and having higher $\Sigma_{\rm SFR}$ than 1-component galaxies at the same \sfr{\re}; they also exhibit higher Gini coefficients.

As shown in Fig.~\ref{fig: SFMS/SFR surface density MS}, there is some overlap between the 1- and multi-component galaxies on the SFMS and the $\Sigma_{\rm SFR}$ main sequence. 
To determine what distinguishes the two samples in the overlapping regions of these parameter spaces, in the 4th column of Fig.~\ref{fig: matched sample analysis - punnet square} we match in both \mstar{} and \sfr{\re}, which yields similar results to matching in \sfr{\re} alone: the matched multi-component galaxies have significantly higher \sfrdens{\re} and \sfrdens{{\rm 3\,kpc}}, and are more compact than the 1-component galaxies. 
Note that we do not simultaneously match in observational parameters as the resulting sample sizes become too small to perform statistical analysis; however, this is unlikely to bias our results, having just shown that the multi-component galaxies exhibit elevated $\Sigma_{\rm SFR}$ when matched in \mstar\ and \sfr{\re} even when controlling for \ha{} A/N, angular scale and inclination. 
In the 5th column of Fig.~\ref{fig: matched sample analysis - punnet square}, we match in both \mstar{} and \sfrdens{\re}. In this case, the matched 1-component galaxies have similar \sfr{\re}, suggesting that $\Sigma_{\rm SFR}$ is more fundamental than SFR in determining the presence of complex emission line profiles in these galaxies.	

We now match in \sfrdens{\re} (columns 9--12 of Fig.~\ref{fig: matched sample analysis - violin plots}). 
Note that the nearly disjoint distributions of the 1- and multi-component galaxies in \sfrdens{\re} (as shown in Fig.~\ref{fig: sample histograms (observational)}) results in a relatively small sample size of 18 when matching in this quantity, and that the matched 1-component galaxies may not be representative of the overall sample.
None the less, at a fixed \sfrdens{\re}, the multi-component galaxies are significantly larger, more massive and have higher \sfr{\re} than the 1-component galaxies; these trends persist when controlling for \ha{} A/N, angular scale and inclination.
Although the multi-component galaxies have a median \sfrdens{{\rm 3\,kpc}} over 0.5\,dex higher than those of the matched 1-component galaxies, this may not be representative of the true $\Sigma_{\rm SFR}$ within the galaxy itself due to their small size. For this reason, we do not match in \sfrdens{{\rm 3\,kpc}}. 
Rather, we propose that these high-$\Sigma_{\rm SFR}$ 1-component systems may harbour the same underlying physical phenomena that manifest as complex emission line profiles in more massive systems (e.g., winds) but that it may not be possible to detect them at the spectral resolution of SAMI; these high-$\Sigma_{\rm SFR}$ 1-component galaxies are discussed further in Section~\ref{subsubsec: high-SFR surface density 1-component galaxies}.

As is clear from Fig.~\ref{fig: matched sample analysis - violin plots}, the multi-component galaxies exhibit significantly higher $\Sigma_{\rm SFR}$ than the 1-component values regardless of which quantity is held constant, even when controlling for observational biases. We therefore conclude that $\Sigma_{\rm SFR}$ is the most important physical parameter in determining the presence of complex emission line profiles in star-forming galaxies. 
However, \ha{} A/N, spectral resolution and inclination effects may play a secondary role; these are discussed further in Sections~\ref{subsubsec: high-SFR surface density 1-component galaxies}, \ref{subsubsec: Spectral resolution limitations} and \ref{subsubsec: Inclincation effects}.

\subsubsection{High-\sfrdens{\re} 1-component galaxies}\label{subsubsec: high-SFR surface density 1-component galaxies}

\begin{figure*}
	\includegraphics[height=0.87\textheight]{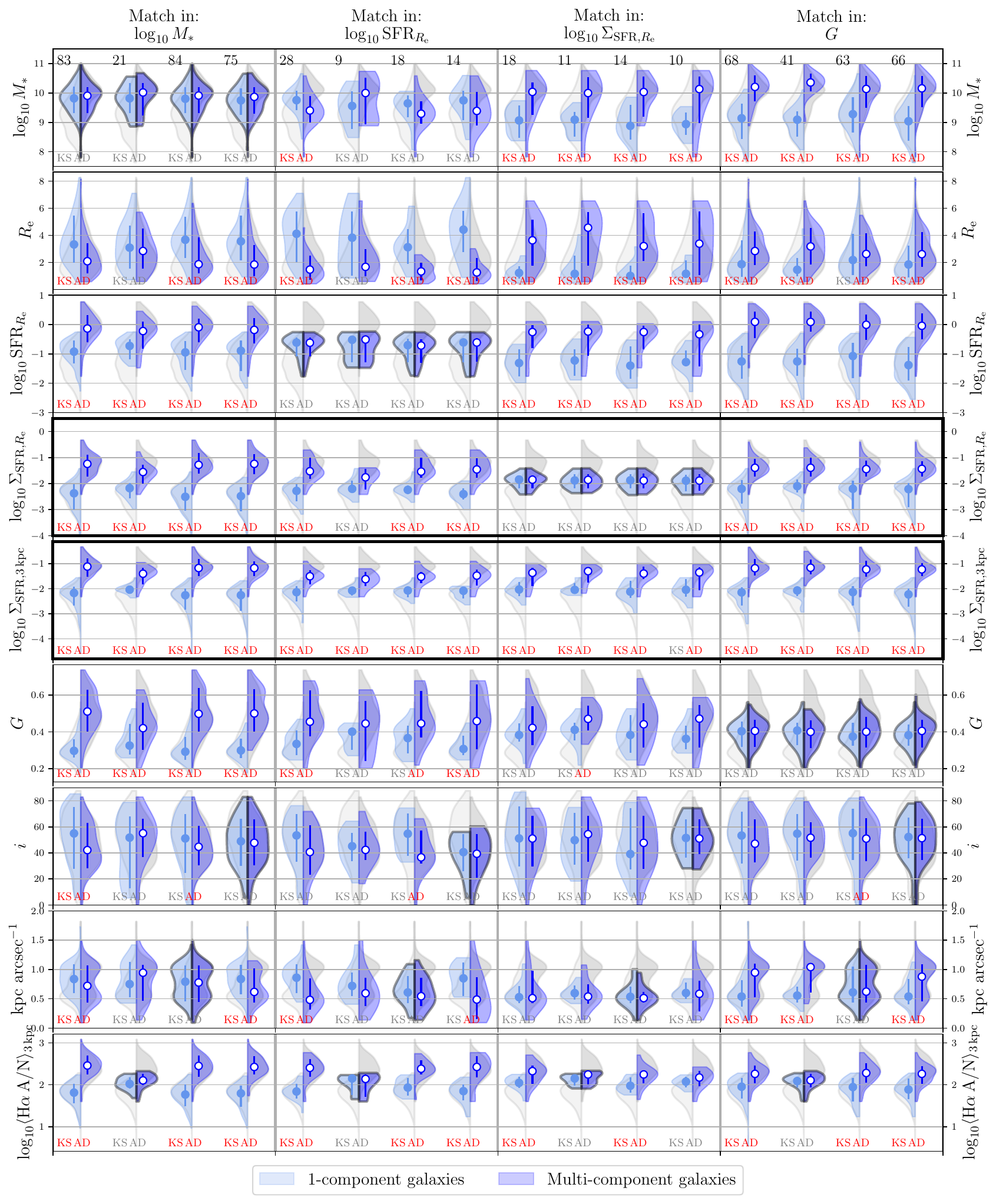}
	\caption{
			Violin plots showing KDEs representing the distributions in various properties for the 1-component (light blue, left) and multi-component (dark blue, right) matched samples. The full distributions are also shown in grey for reference. 
			Each row shows the distributions in the matched samples in various quantities, and each column shows the distributions matched in the quantity indicated by thick grey outlines around violin plots. 
			The figure is divided into four sub-groups representing samples matched in the quantity shown at the top of the figure: \mstar\, \sfr{\re}, \sfrdens{\re} and Gini coefficient $G$. 
			Within each sub-group, the leftmost column of violin plots represents samples matched only in that quantity, and the other three columns represent samples matched in that quantity simultaneously with (from left to right) mean \ha{} A/N within 3\,kpc, angular scale, and inclination.
			In each violin plot, the circular points and error bars show the mean and interquartile ranges (IQRs) of the 1- and multi-component matched samples, and the number of galaxies in each matched sample is shown in the upper left.
			The text at the bottom of each plot represents the result of KS and AD 2-sample tests; text in red represents a $p$-value less than 0.05, indicating a statistically significant likelihood that the distributions of the matched sample in that quantity are not drawn from the same parent distribution.
	} 
	\label{fig: matched sample analysis - violin plots}
\end{figure*}

As shown in columns 9--12 of Fig.~\ref{fig: matched sample analysis - violin plots}, our sample contains a small population of 1-component galaxies with high $\Sigma_{\rm SFR}$.

Matching in \sfrdens{\re} alone yields 18 compact, low-mass 1-component galaxies with high \sfrdens{\re} that have significantly lower \sfrdens{{\rm 3\,kpc}} than the matched multi-component galaxies. This could indicate that \sfrdens{{\rm 3\,kpc}} is fundamentally more important than \sfrdens{\re}; however, it is also possible that \sfrdens{{\rm 3\,kpc}} is not a reliable estimate of the true $\Sigma_{\rm SFR}$ within these galaxies due to their small sizes, as stated in Section~\ref{subsec: General properties of 1- and multi-component galaxies}.
However, matching in both \mstar{} and \sfrdens{\re} (5th column of Fig.~\ref{fig: matched sample analysis - punnet square}) reveals a sample of 1-component galaxies with similar masses and sizes to the matched multi-component galaxies. 
Notably, these galaxies have lower \sfrdens{{\rm 3\,kpc}} by around 0.4\,dex but have similar sizes to the matched multi-component galaxies, suggesting the offset in \sfrdens{{\rm 3\,kpc}} is not an artefact of small galaxy sizes.

The lack of complex emission line profiles in the high-\sfrdens{\re} 1-component systems may be a spectral resolution effect: as discussed in Section~\ref{subsubsec: Spectral resolution limitations}, we may be unable to resolve multiple emission line components in low-mass systems at the spectral resolution of the SAMI instrument. Higher-resolution observations may therefore reveal complex emission line components in these unusual objects.

\section{Correlation analysis}
\label{sec: correlation analysis}

\begin{figure}
	\centering
	\includegraphics[width=\linewidth]{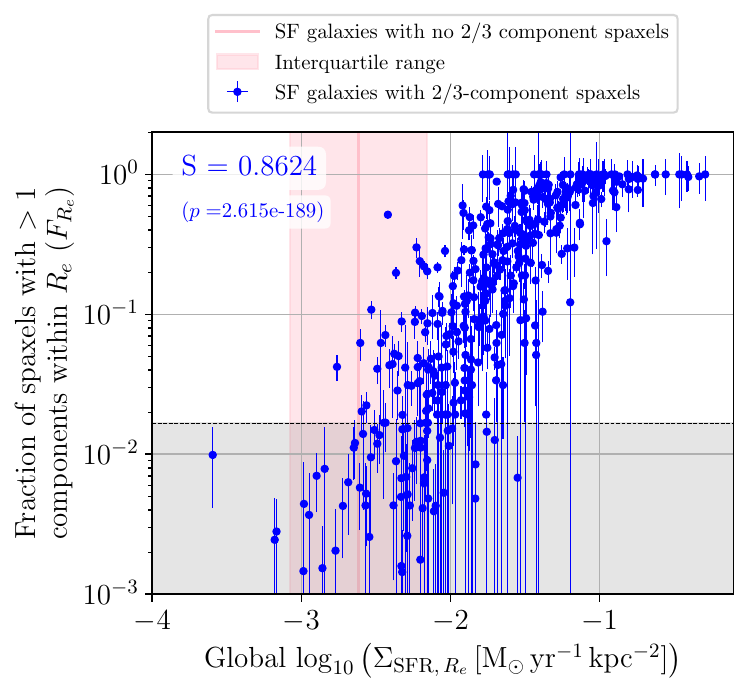}
	\caption{
		Fraction of spaxels within 1\re{} containing multiple components ($F_{\re}$) as a function of \sfrdens{\re}. Star-forming galaxies with $F_{\re} > 0$ are shown in blue, where the vertical error bars represent $1\sigma$ Poisson errors. The pink vertical line and shaded region indicate the mean and interquartile ranges of global \sfrdens{\re} of galaxies containing only 1-component spaxels. Spearman's rank correlation coefficients and associated $p$-values computed for star-forming galaxies with $F_{\re} > 0$ are shown in the upper left-hand corner. 
		The horizontal dashed line represents the minimum reliable value for $F_{\re}$, corresponding to one resolution element containing multiple components as discussed in Section~\ref{sec: local vs. global effects}.
		Values below the horizontal dashed may be unreliable due to resolution effects.
		A version of this figure made using measurements from a 3\,kpc aperture is shown in Fig.~\ref{fig: F vs. global SFR surface density (3kpc)}.
	}
	\label{fig: F vs. global SFR surface density}
\end{figure}

In Fig.~\ref{fig: F vs. global SFR surface density}, we plot $F_{\re}$, the fraction of spaxels within 1\re{} containing two or three components within an individual galaxy, as a function of global \sfrdens{\re} for star-forming SAMI galaxies. 
Note that although the pixel size of SAMI is 0.5'', the effective number of resolution elements across the SAMI FoV is approximately 60, meaning that values of $F_{\re} < 1 / 60$ may represent artefacts, which may explain the increase in scatter at low values of $F_{\re}$; this threshold is indicated by the shaded grey region. 
There is a strong and statistically significant correlation between $F_{\re}$ and \sfrdens{\re} as indicated by the Spearman's rank correlation coefficient of 0.862 ($p \ll 0.05$). Moreover, galaxies with $F_{\re} = 0$ have markedly lower global $\Sigma_{\rm SFR}$ than those with $F_{\re} > 0$, as indicated by the pink shaded region. A similarly strong correlation is observed between $F_{\rm 3\,kpc}$ and \sfrdens{{\rm 3\,kpc}} (Fig.~\ref{fig: F vs. global SFR surface density (3kpc)}).
This trend is consistent with the finding from our matched sample analysis that $\Sigma_{\rm SFR}$ is the parameter that is most strongly correlated with the presence of complex emission line profiles; however, Figs.~\ref{fig: F vs. global SFR surface density} and ~\ref{fig: F vs. global SFR surface density (3kpc)} also reveal that the association between $F$ and $\Sigma_{\rm SFR}$ is continuous in nature, albeit with significant scatter.

For completeness, correlation strengths between $F_{\re}$ and parameters other than $\Sigma_{\rm SFR}$ are shown in Table~\ref{tab: correlation coefficients}; amongst the parameters considered, \sfrdens{\re} indeed shows the strongest correlation with $F_{\re}$ (with a strength of 0.862), followed by \sfrdens{3\,\kpc} (0.858). However, there is also a strong correlation between $F_{\re}$ and the mean \ha{} A/N measured within a 3\,kpc aperture (0.729).

To determine whether A/N is responsible for the correlation between $\Sigma_{\rm SFR}$ and $F_{\re}$, we compute partial correlation coefficients (PCCs) in order to examine the influence of confounding variables such as A/N and angular scale upon this correlation. These values are shown in Table~\ref{tab: partial correlation coefficients}, and were calculated using the \texttt{partial\_corr} function from the \textsc{python} package \textsc{pingouin}~\citep{Vallat2018}.
In all cases, $p \ll 0.05$, indicating statistically a significant correlation between $\Sigma_{\rm SFR}$ and $F_{\re}$ when controlling for all confounding variables considered here.
Of all variables, \ha{} A/N has the strongest impact on the correlation between $F_{\re}$ and \sfrdens{\re}, confirming that A/N does play a role in determining the fraction of spaxels within any given galaxy that exhibit multiple components. However, when controlling for this effect, the correlation between $F_{\re}$ and \sfrdens{\re} remains strong at approximately $0.69$, indicating that indeed $\Sigma_{\rm SFR}$ is the variable that is most strongly associated with the presence of complex emission line profiles, and that the higher A/N associated with multi-component spaxels is most likely due to elevated \ha{} fluxes resulting from more concentrated star formation.
This is consistent with our matched sample analysis, which showed that multi-component galaxies exhibit elevated $\Sigma_{\rm SFR}$ compared to 1-component galaxies when matched in A/N (Section~\ref{sec: Matched sample analysis}). 
We therefore conclude that the correlation shown in Fig.~\ref{fig: F vs. global SFR surface density} is unlikely to be driven purely by A/N.

\begin{table}
	\caption{
		Spearman's rank-order correlation coefficients $S$ and associated $p$-values computed for correlations between the fraction of multi-component spaxels ($F_{\re}$) and the quantities given in the leftmost column for the 655 star-forming galaxies in our sample. The strongest correlations, highlighted in bold, are between $F_{\re}$ and \sfrdens{\re} and \sfrdens{{\rm 3\,kpc}}. To enable a fair comparison, for quantities indicated by the daggers, the correlation was computed for $F_{\rm 3\,kpc}$ instead of $F_{\re}$.
	}
	\centering
	\begin{tabular}{ccc}
		\hline
		\textbf{Parameter} & \textbf{Correlation coefficient} & \textbf{$p$-value} \\
		\hline
		$\log_{10}(M_*\,[\rm M_\odot])$ &   $0.423$   &   $8.646 \times 10^{-29}$\\
		$\log_{10}$ \sfr{\re}   &   $0.718$   &   $1.412 \times 10^{-101}$\\
		$\log_{10}$ \sfrdens{\re}   &   \textbf{$0.862$}  &   $2.615 \times 10^{-189}$\\
		$\log_{10}$ \sfrdens{3\,\kpc}   &   \textbf{$0.858$}  &   $2.877 \times 10^{-183}$\\
		$\log_{10}$ s\sfr{\re}  &   $0.549$   &   $7.941 \times 10^{-51}$\\
		$\log_{10}$ s\sfr{3\,\kpc}  &   $0.404$   &   $7.664 \times 10^{-26}$\\
		$R_e$ [kpc] &   $-0.049$  &   $0.2165$\\
		$\log_{10}(M_*/R_e\,[\rm M_\odot\,kpc^{-1}])$   &   $0.590$   &   $2.284 \times 10^{-60}$\\
		$\log_{10}(M_*/R_e^2\,[\rm M_\odot\,kpc^{-2}])$ &   $0.675$   &   $6.025 \times 10^{-85}$\\
		$g - i$ colour  &   $0.259$   &   $4.151 \times 10^{-11}$\\
		$G$ &   $0.644$   &   $9.921 \times 10^{-76}$\\
		$\log_{10} \langle {\rm H}\upalpha \, {\rm A/N} \rangle_{\rm 3\,kpc} $  &   \textbf{$0.729$}  &   $2.368 \times 10^{-106}$\\
		$z$ &   0.258   &   $4.265 \times 10^{-11}$\\
		$\rm kpc \, arcsec^{-1}$    &   $0.258$   &   $4.265 \times 10^{-11}$\\
		$i$ [degrees]   &   $-0.126$  &  $0.001479$ \\
		\hline 
	\end{tabular}
	\label{tab: correlation coefficients}
\end{table}

\begin{table}
	\caption{Partial Spearman's rank correlation coefficients between $\log_{10} \sfrdens{\re}$ and $F_{\re}$ when controlling for various parameters. Corresponding $p$-values are not shown as they are $\ll 0.05$ in all cases.}
	\centering
	\vspace{0.5cm}
	\begin{tabular}{cc}
		\hline
		\textbf{Parameter} & \textbf{Partial correlation coefficient} \\
		\hline
        $\log_{10}(M_*\,[\rm M_\odot])$                    & 0.845\\
		$\log_{10}$ \sfr{\re} 							   & 0.731\\
		$\log_{10}$ s\sfr{\re}  						   & 0.795\\
		$\log_{10}$ s\sfr{3\,\kpc}          			   & 0.837\\
		$R_e$ [kpc]                                        & 0.878\\
		$\log_{10}(M_*/R_e\,[\rm M_\odot\,kpc^{-1}])$      & 0.794\\
		$\log_{10}(M_*/R_e^2\,[\rm M_\odot\,kpc^{-2}])$    & 0.726\\
		$g - i$ colour                                     & 0.852\\
		$G$                                                & 0.758\\
		$\log_{10} \langle {\rm H}\upalpha \, {\rm A/N} \rangle_{\rm 3\,kpc} $    & \textbf{0.686}\\
		$z$                                                & 0.853\\
		$\rm kpc \, arcsec^{-1}$                           & 0.853\\
		$i$ [degrees]                          			   & 0.868\\
		\hline
	\end{tabular}
	\label{tab: partial correlation coefficients}
\end{table}

\section{Is the presence of multiple emission line components driven by star formation on local or global scales?}
\label{sec: local vs. global effects}

As shown in Sections~\ref{sec: Matched sample analysis} and \ref{sec: correlation analysis}, galaxies containing complex emission line profiles are associated with higher \sfrdens{\re} and \sfrdens{{\rm 3\,kpc}}, i.e., \textit{global} $\Sigma_{\rm SFR}$, even after controlling for both physical and observational effects.
However, because the global $\Sigma_{\rm SFR}$ is a measure of the average $\Sigma_{\rm SFR}$ within individual spaxels, it is unclear whether the presence of complex emission line profiles---and therefore winds, outflows and/or thick ionised gas disks---are associated with star formation on local, or global (i.e., \textit{galaxy-wide}) scales.
We now leverage the spatially resolved nature of the SAMI data set to determine whether star formation on local or global scales is the underlying driver of the presence of multiple emission line components.

\begin{figure}
	\centering
	\includegraphics[width=\linewidth]{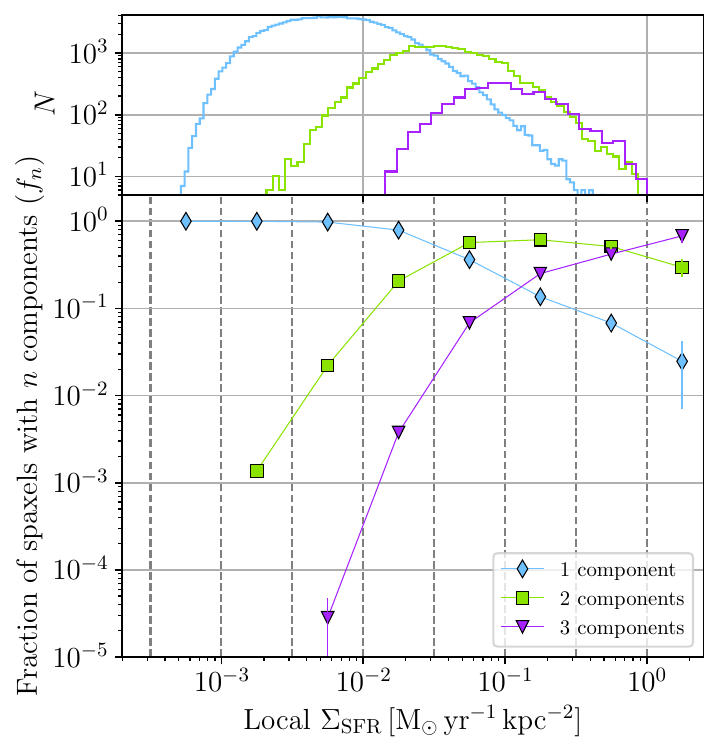}
	\caption{
		Fraction $f_n$ of star-forming spaxels in the SAMI Galaxy Survey containing 1 ($f_1$, blue diamonds), 2 ($f_2$, green squares) and 3 ($f_3$, purple triangles) emission line components as a function of spaxel-scale (i.e., \textit{local}) $\Sigma_{\rm SFR}$. $1\sigma$ Poisson errors in $f_n$ are indicated by the vertical black error bars. The log-scaled histogram in the top panel shows the distribution of 1, 2 and 3-component spaxels in $\Sigma_{\rm SFR}$.
		Vertical dashed lines indicate bin edges.
	}
	\label{fig: f vs. local SFR surface density}
\end{figure}

In Fig.~\ref{fig: f vs. local SFR surface density} we show $f_n$, or the fraction of star-forming spaxels in the full SAMI data set containing $n$ components, as a function of \textit{local} $\Sigma_{\rm SFR}$ (i.e., that measured within individual spaxels), representing the spatially-resolved analogue of Fig.~\ref{fig: F vs. global SFR surface density}. 
Before continuing, we note that although the spaxel size of SAMI is 0.5'', the effective spatial resolution is approximately 2'', corresponding to a median physical scale of approximately 1\,kpc for the galaxies in our sample (Fig.~\ref{fig: sample histograms}).
$f_1$ and $f_2$ increase smoothly as a function of $\Sigma_{\rm SFR}$ such that 2- and 3-component spaxels begin to dominate at $\Sigma_{\rm SFR} \gtrsim 10^{-1.5}\,\msol\,\yr\,\kpc^{-2}$.  
Together, Figs.~\ref{fig: F vs. global SFR surface density} and \ref{fig: f vs. local SFR surface density} indicate that both local and global effects play a role in determining whether or not complex emission line profiles are present. 

\begin{figure}
	\centering
	\includegraphics[width=\columnwidth]{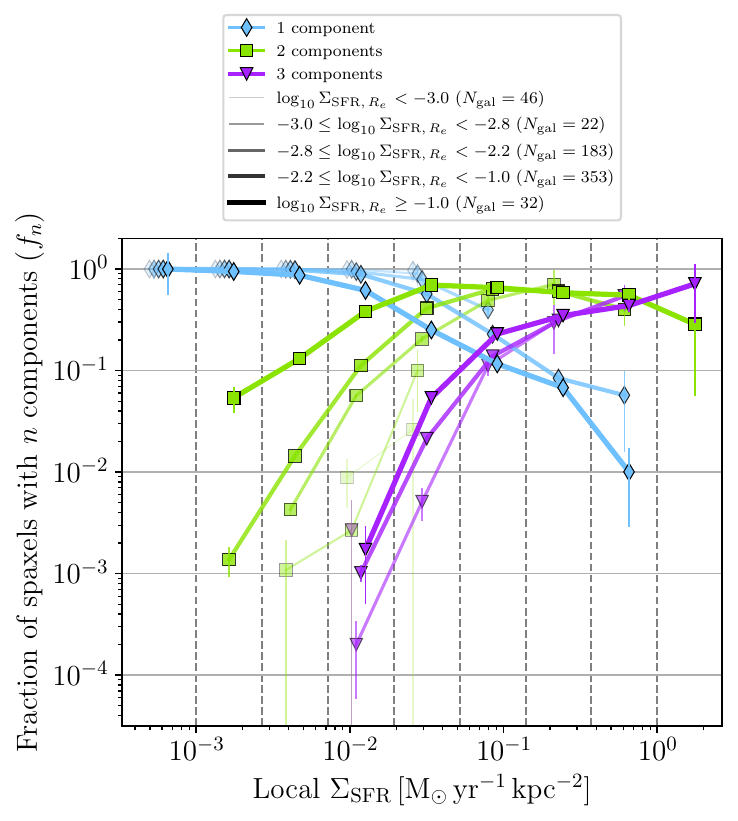}
	\caption{
		Same as Fig.~\ref{fig: f vs. local SFR surface density}, where spaxels have been binned by the \sfrdens{\re} (i.e., \textit{global} $\Sigma_{\rm SFR}$) of their host galaxies. Thicker lines indicate spaxels residing in galaxies with higher global \sfrdens{\re}. 
		Within each bin of local $\Sigma_{\rm SFR}$, the points have been slightly shifted in $x$ for clarity.
		In the legend, $N_{\rm gal}$ refers to the number of galaxies in each $\Sigma_{\rm SFR,\,\re}$ bin.
		Vertical dashed lines indicate bin edges.
		The likelihood of a spaxel exhibiting complex emission line profiles therefore depends not only upon the local $\Sigma_{\rm SFR}$, but on the global $\Sigma_{\rm SFR}$ of its host galaxy.
	}
	\label{fig: f vs. local SFR surface density (binned by SFR surface density)}
\end{figure}

To further investigate the relationship between $f_n$ and the roles of local versus global star formation, in Fig.~\ref{fig: f vs. local SFR surface density (binned by SFR surface density)} we plot $f_n$ as a function of local $\Sigma_{\rm SFR}$ in our sample of star-forming galaxies, but in bins of host galaxy \sfrdens{\re}. At a fixed local $\Sigma_{\rm SFR}$, spaxels residing in galaxies with higher global $\Sigma_{\rm SFR}$ are much more likely to exhibit multiple emission line components. This effect is even stronger when \sfrdens{{\rm 3\,kpc}} is used as a measure of the global $\Sigma_{\rm SFR}$ (Fig.~\ref{fig: f vs. local SFR surface density (binned by SFR surface density (3kpc))}). 
The likelihood of detecting multiple emission line components in any given spaxel is therefore affected not only by the local $\Sigma_{\rm SFR}$, but by star formation occurring on scales larger than the spatial resolution of SAMI.
This illustrates that care must be taken when studying gas kinematics within individual spaxels when divorced from their host galaxy properties.

To investigate the influence of observational biases on this result, in Figs.~\ref{fig: f vs. local SFR surface density (binned by inclination)} and ~\ref{fig: f vs. local SFR surface density (binned by angular scale)} we show $f_n$ vs. $\Sigma_{\rm SFR}$ in spaxels binned by their host galaxy's inclination and angular scale. At a fixed $\Sigma_{\rm SFR}$, the relative enhancement in $f_2$ and $f_3$ in galaxies at different angular scales and inclinations tends to be smaller than that in galaxies with different global $\Sigma_{\rm SFR}$, indicating that these observational biases do not primarily drive the splitting between low- and high-$\Sigma_{\rm SFR}$ galaxies shown in Fig.~\ref{fig: f vs. local SFR surface density (binned by SFR surface density)}. 
We do note, however, that spaxels in more face-on galaxies are more likely to exhibit multiple emission line components than those in edge-on systems; this is most likely due to off-planar emission being present as a single emission line component when viewed in a near-edge-on system. The effects of inclination are discussed further in Section~\ref{subsubsec: Inclincation effects}.

Under certain circumstances, beam smearing can produce artificial broad emission line components due to smearing of a steep velocity gradient by seeing effects~\citep{Marasco2019,Concas2022}. SFR profiles tend to peak in the nuclei of galaxies, meaning that spaxels with high $\Sigma_{\rm SFR}$ are more likely to be affected by beam smearing. 
To determine whether the increase in $f_n$ with $\Sigma_{\rm SFR}$ could be partially due to artificial broad line components caused by beam smearing, in Fig.~\ref{fig: f vs. local SFR surface density (binned by spaxel radius)} we plot $f_n$ as a function of local $\Sigma_{\rm SFR}$ binned by projected spaxel radius. At a fixed local $\Sigma_{\rm SFR}$, spaxels located within the seeing disc (approximately 2'') are marginally more likely to exhibit multiple emission line components, but this effect is far smaller than the offsets seen between spaxels in galaxies with different global $\Sigma_{\rm SFR}$ in Fig.~\ref{fig: f vs. local SFR surface density (binned by SFR surface density)}.
We therefore conclude that the correlation between global $\Sigma_{\rm SFR}$ and the presence of complex emission line profiles has a physical origin, and that star formation taking place on global scales can affect the kinematics of gas on smaller scales as seen in IFS observations.

\subsection{How important is clumpiness?}\label{subsec: How important is clumpiness?}

To further investigate the contributions from $\Sigma_{\rm SFR}$ on local and global scales, we now consider the clumpiness of the star formation distribution within individual galaxies. 
As shown in Fig.~\ref{fig: F vs. global SFR surface density}, there is considerable vertical scatter in $F_{\re}$ versus \sfrdens{\re}.
If star formation on local, rather than global, scales is the fundamental driver of multiple emission line components, then galaxies with similar global $\Sigma_{\rm SFR}$, but clumpier SFR distributions (i.e., containing spaxels with higher local $\Sigma_{\rm SFR}$), should have a larger $F$, whereas if global $\Sigma_{\rm SFR}$ is more important, then all galaxies should have similar values of $F$ regardless of local variations in $\Sigma_{\rm SFR}$. 
That is, we would expect clumpiness to contribute to the vertical scatter in the relation shown in Fig.~\ref{fig: F vs. global SFR surface density}.

\begin{figure}
	\centering
	\includegraphics[width=1\linewidth]{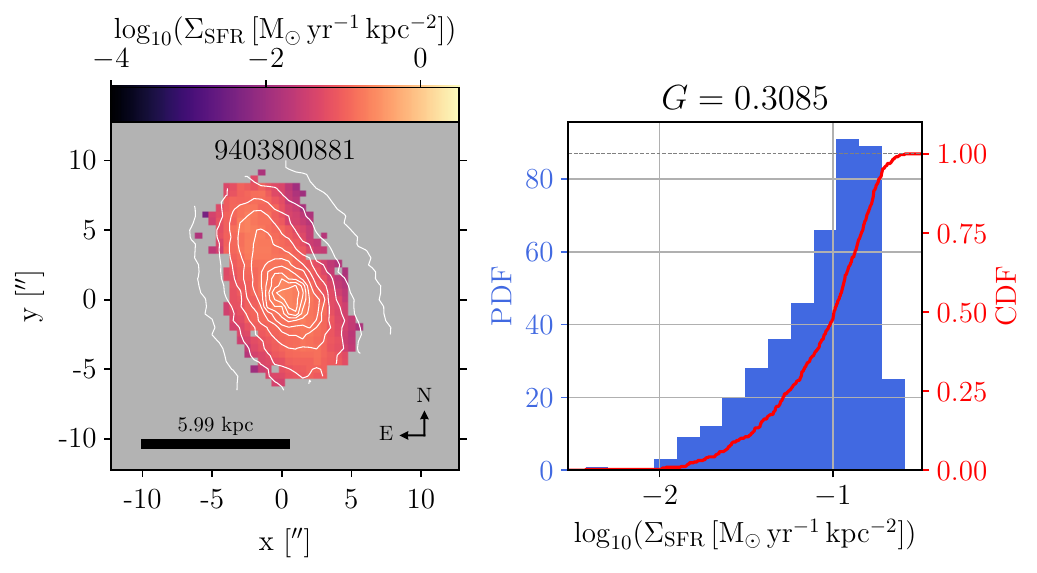}\\(a)\\
	\includegraphics[width=1\linewidth]{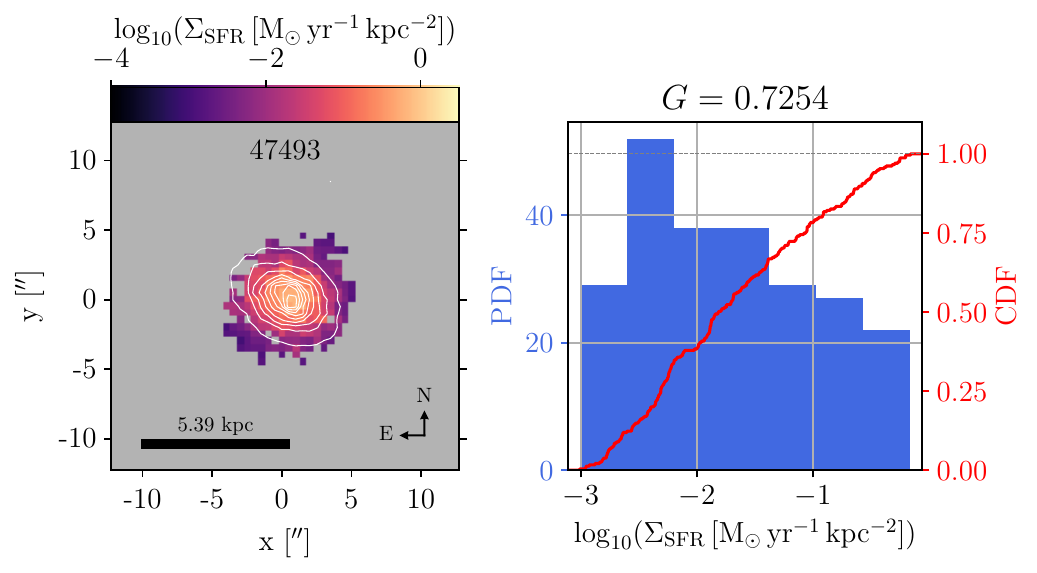}\\(b)
	\caption{Maps, PDFs and cumulative distribution functions (CDFs) in $\log_{10} \Sigma_{\rm SFR}$ for (a) galaxy 9403800881, which has a low Gini coefficient of $0.3085$, and (b) 47493, which has a high Gini coefficient of $0.7254$.}
	\label{fig: Gini coefficient examples}
\end{figure}

To quantify the clumpiness of star formation we employ the Gini coefficient~\citep{Gini1936}, a parameter originally developed to quantify wealth inequality that has previously been applied to parametrise the clumpiness of galaxy morphologies by \citet{Lotz2004} and \citet{Bloom2017}. Gini coefficients for each galaxy were computed from the SFR maps using Eqn.~\ref{eq: Gini coefficient}. Figure~\ref{fig: Gini coefficient examples} shows $\Sigma_{\rm SFR}$ maps and corresponding Gini coefficients for two SAMI galaxies; galaxy 9403800881 has a much smoother $\Sigma_{\rm SFR}$ distribution and therefore a much lower Gini coefficient ($G=0.3085$) than 47493 ($G=0.7254$), which has centrally-concentrated star formation.

As shown in Fig.~\ref{fig: sample histograms}, multi-component galaxies have higher average $G$ than 1-component galaxies, implying clumpier and/or more centrally concentrated star formation, although there is substantial overlap between the two distributions. 
multi-component galaxies also have higher median $G$ when matched in stellar mass, SFR, and other quantities as shown in Fig.~\ref{fig: matched sample analysis - violin plots}.
However, as shown in Table~\ref{tab: partial correlation coefficients}, the correlation between $F_{\re}$ and \sfrdens{\re} remains very strong at 0.7578 when controlling for $G$, suggesting that at the spectral and spatial resolution of SAMI, global $\Sigma_{\rm SFR}$ is more important than local variations in $\Sigma_{\rm SFR}$ within galaxies in determining the presence of complex emission line profiles.


\section{Discussion}\label{sec: Discussion}

\subsection{Physical interpretation}\label{subsec: Physical interpretation}

In Section~\ref{sec: Matched sample analysis}, we showed that multi-component galaxies have systematically higher \sfrdens{\re} than the 1-component galaxies.
We also determined in Section~\ref{sec: local vs. global effects} that the presence of multiple emission line components is driven by $\Sigma_{\rm SFR}$ on both global and local scales. Observational biases do not explain these results, indicating an underlying physical mechanism.

Complex emission line profiles in star-forming galaxies are most often interpreted as winds~\citep[e.g.,][]{McKeith1995,Westmoquette2009a,Ho2014} or thick ionised gas disks~\citep[also referred to as extraplanar gas, or DIG; e.g.,][]{denBrok2020,Belfiore2022}; both of these phenomena can be interpreted as manifestations of galactic fountains~\citep{Fraternali&Binney2006,Barnabe2006,Marinacci2010a}.
Although some SAMI galaxies exhibit complex gas kinematics due to processes such as mergers or ram-pressure stripping~(e.g., Quattropani et al., \textit{in prep.}), visual inspection of SDSS imaging indicates that only approximately 10\,per\,cent of our multi-component galaxies are undergoing a major merger.
We therefore assume that the presence of complex emission line profiles in the star-forming galaxies in our sample indicates the presence of galactic fountains in the form of winds or thick disks.
Detailed kinematic analysis of the multi-component galaxies, including consideration of the differences between galaxies harbouring winds and those with thick disks, is deferred to a future work.

Our finding that $\Sigma_{\rm SFR}$ is the parameter most strongly associated with complex emission line profiles and winds is consistent with previous findings using SAMI data. 
\citet{Ho2016a} found that edge-on systems with winds have systematically higher $\Sigma_{\rm SFR}$ than those without. 
\citet{Tescari2018} additionally found that extraplanar gas in simulated galaxies with higher $\Sigma_{\rm SFR}$ tend to exhibit high-$\sigma$ tails, which the authors interpreted as increased outflow activity due to star formation. More recently, in a sample of star-forming SAMI galaxies, \citet{Varidel2020} found the vertical gas velocity dispersion correlates more strongly with $\Sigma_{\rm SFR}$ than with SFR. 

Recent studies using MaNGA~\citep{Bundy2015} have also found strong links between $\Sigma_{\rm SFR}$ and winds. 
\citet{Avery2021} found that galaxies with star formation-driven outflows (as identified by the presence of broad \ha{} components in aperture spectra) tend to have overall $\Sigma_{\rm SFR} \gtrsim 10^{-2}\,\msol\,\yr^{-1}\,\kpc^{-2}$. Earlier, \citet{Roberts-Borsani2020a} found that neutral outflows (as traced by interstellar Na\,D absorption) are only detected in galaxies with $\Sigma_{\rm SFR} \gtrsim 10^{-2}\,\msol\,\yr^{-1}\,\kpc^{-2}$. These thresholds are remarkably similar to that we find between the 1- and multi-component galaxies (Fig.~\ref{fig: sample histograms}).

Similar trends have been found between the presence of thick ionised gas disks and $\Sigma_{\rm SFR}$. \citet{Rossa&Dettmar2003a} detected extraplanar gas only in galaxies above a certain $\Sigma_{\rm SFR}$ threshold, and using a small sample of galaxies, \citet{Rueff2013} found that DIG becomes brighter and more filamentary as $\Sigma_{\rm SFR}$ increases. 
More recently, \citet{Belfiore2022} found that the velocity dispersions of DIG-dominated spaxels (identified by enhanced forbidden line emission) only become significantly larger than the velocity dispersions in \hii{} region-dominated spaxels in galaxies with high global SFRs. This suggests that at sufficiently low SFRs, it may not be possible to resolve DIG from \hii{} region emission using our spectral decomposition technique. Together with the correlation between stellar mass and SFR (Fig.~\ref{fig: SFMS/SFR surface density MS}), this may explain why there are few multi-component galaxies with $\mstar \lesssim 10^{9}\,\msol$. These and other spectral resolution considerations are discussed in Section~\ref{subsubsec: Spectral resolution limitations}.

It is therefore possible that these thick gas disks may be present in most of the star-forming galaxies in our sample, but that they are only detectable with SAMI above a certain $\Sigma_{\rm SFR}$ and/or stellar mass threshold. 
Unresolved thick gas disks could explain the population of high-$\Sigma_{\rm SFR}$ 1-component galaxies as discussed in Section~\ref{subsubsec: high-SFR surface density 1-component galaxies}: as shown in Fig.~\ref{fig: matched sample analysis - violin plots}, these systems have low stellar masses and low SFRs, but are very compact and so have high \sfrdens{\re}. Confirming this scenario would require higher spectral resolution observations.

As shown in Fig.~\ref{fig: f vs. local SFR surface density (binned by SFR surface density)}, spaxels residing in galaxies with higher \sfrdens{\re} are more likely to exhibit multiple emission line components regardless of their local $\Sigma_{\rm SFR}$. 
An analogous effect was found by \citet{Law2022}, who found global SFR to correlate more strongly with the \ha{} velocity dispersion than the local $\Sigma_{\rm SFR}$.
This trend can be explained by projection effects. 
A hot wind originating from a high-$\Sigma_{\rm SFR}$ region of a galaxy will accelerate cloud fragments from the disk outwards in a spherical fashion~\citep{Cooper2008,Cooper2009}, forming dramatic biconical structures such as that observed in M82~\citep{Lynds&Sandage1963,Bland&Tully1988,Shopbell&Bland-Hawthorn1998}. Due to orientation effects, \ha{}-emitting filaments in the wind may therefore intersect lines of sight towards lower-$\Sigma_{\rm SFR}$ regions of the galaxy that are not involved in launching the wind; this would also apply to gas returning to the disk via a galactic fountain, in the process forming a thick ionised gas disk.
Evidence for these LoS effects has been observed in the Large Magellanic Cloud, in which highly ionised winds, traced by broad \forb{O}{vi} absorption, can be seen along lines of sight towards both locations of active star-formation and quiescent regions~\citep{Howk2002,Barger2016}. 
Our results suggest that spectral decomposition may be a viable way to detect this phenomenon in unresolved systems.

\subsection{Observational biases}\label{subsec: Observational biases}

Here, we discuss several sources of observational bias which may be important to consider when carrying out any study relying on multi-component emission line fits.

\subsubsection{Reliability of component identification}\label{subsubsec: Reliability of spaxel classification}
As detailed in \citet{Hampton2017a}, \lzcomp{}, the ANN used to classify spaxels as containing 1, 2 or 3 emission line components, is trained using human observers; as such, its reliability as a function of S/N is unclear. 
Due to our conservative S/N and data quality cuts on the 2nd and 3rd kinematic components within each spaxel (as detailed in Section \ref{subsec: Data quality and S/N cuts}), we assume that our sample of 2- and 3-component spaxels is unlikely to contain misclassified 1-component spaxels.
However, our sample of 1-component spaxels may contain misclassified spaxels that in fact contain 2 (or more) components; here, we perform two tests to determine the degree of contamination within our 1-component spaxel sample. 3-component spaxels are not discussed due to their rarity (representing $< 1$\,per\,cent of all spaxels; see Section~\ref{sec: Defining 1- and multi-component galaxies}).

\begin{figure}
	\centering
	\includegraphics[width=0.9\linewidth]{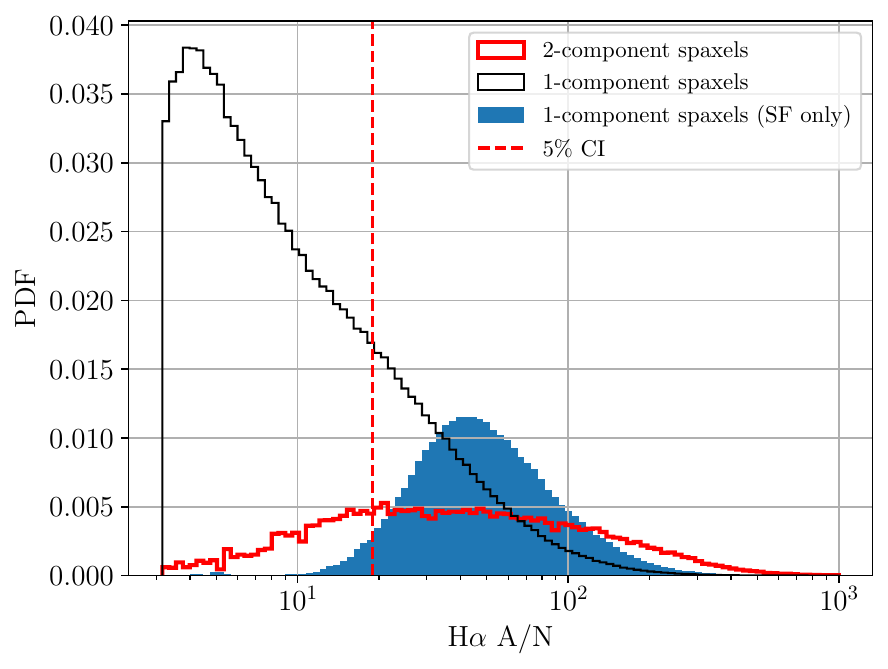}
	\caption{PDFs of \ha{} A/N in all 2-component spaxels (red), all 1-component spaxels (black) and SF 1-component spaxels (blue). The vertical dashed line shows the 5\% confidence interval for the 2-component spaxels.}
	\label{fig: Halpha A/N distributions}
\end{figure}

First, we investigate the impact of A/N on the classification process, where A/N is defined as the peak amplitude of the \ha{} line (measured directly from the spectrum, and therefore not corrected for stellar absorption) relative to the mean continuum level within the rest-frame wavelength range 6500\,Å--6540\,Å divided by the RMS noise within the same range.
We assume that \lzcomp{} begins to misclassify multi-component spaxels as 1-component below some A/N threshold, which we define as the 5\,per\,cent confidence interval of the A/N distribution of the 2-component spaxels that pass our S/N and data quality cuts.
In Fig.~\ref{fig: Halpha A/N distributions} we show distributions in A/N in all 2-component spaxels (red), all 1-component spaxels (black) and SF 1-component spaxels (blue). The vertical dashed line shows the 5\% confidence interval for the 2-component spaxels of approximately 19.
Whilst over 40\% of all 1-component spaxels have A/N below this threshold, fewer than 2\% of SF spaxels fail this criterion. Under the assumption that A/N is an important factor influencing the ANN, this implies that the SF 1-component spaxels, which are the focus of this work, are unlikely to be contaminated by a significant amount of low-S/N 2-component spaxels.

\begin{figure}
		\centering
		\includegraphics[width=0.75\linewidth]{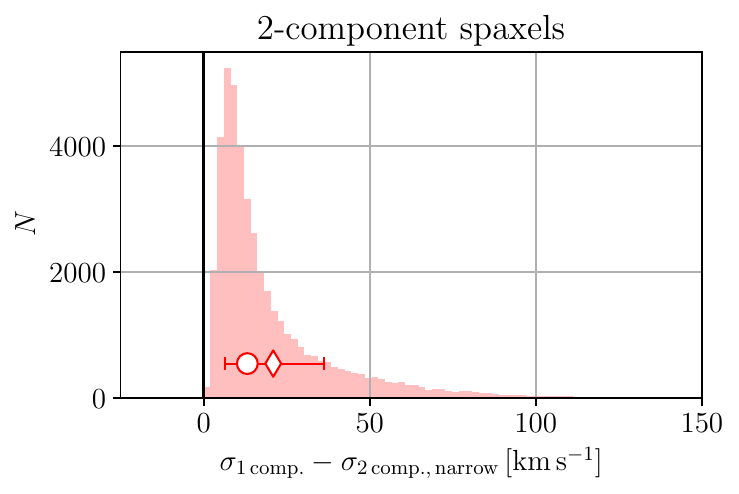}
		\caption{Distribution in the difference between the velocity dispersion of the 1-component fit and the narrow component in the 2-component fit for 2-component spaxels. The median and mean in the distribution are indicated by the white circle and diamond respectively, where the horizontal error bar represents the interquartile range.}
		\label{fig: observational biases: delta sigma_gas}
	\end{figure}

	We now check the reliability of the ANN classification by inspecting the velocity dispersions of components measured within 1- and 2-component spaxels.
	When 2-component spaxels are fitted with a single component, the width of the best-fit Gaussian tends to be broader by an average of around $10\,\kms$ than that of the narrow component in the 2-component fit, as shown in Fig.~\ref{fig: observational biases: delta sigma_gas}.
	We therefore compare the measured velocity dispersions in 1- and 2-component spaxels to estimate the contamination of misclassified multi-component spaxels in our sample of 1-component spaxels. 
	
\begin{figure}
		\centering
		\begin{subfigure}[b]{1.0\linewidth}
			\centering
			\includegraphics[width=\linewidth]{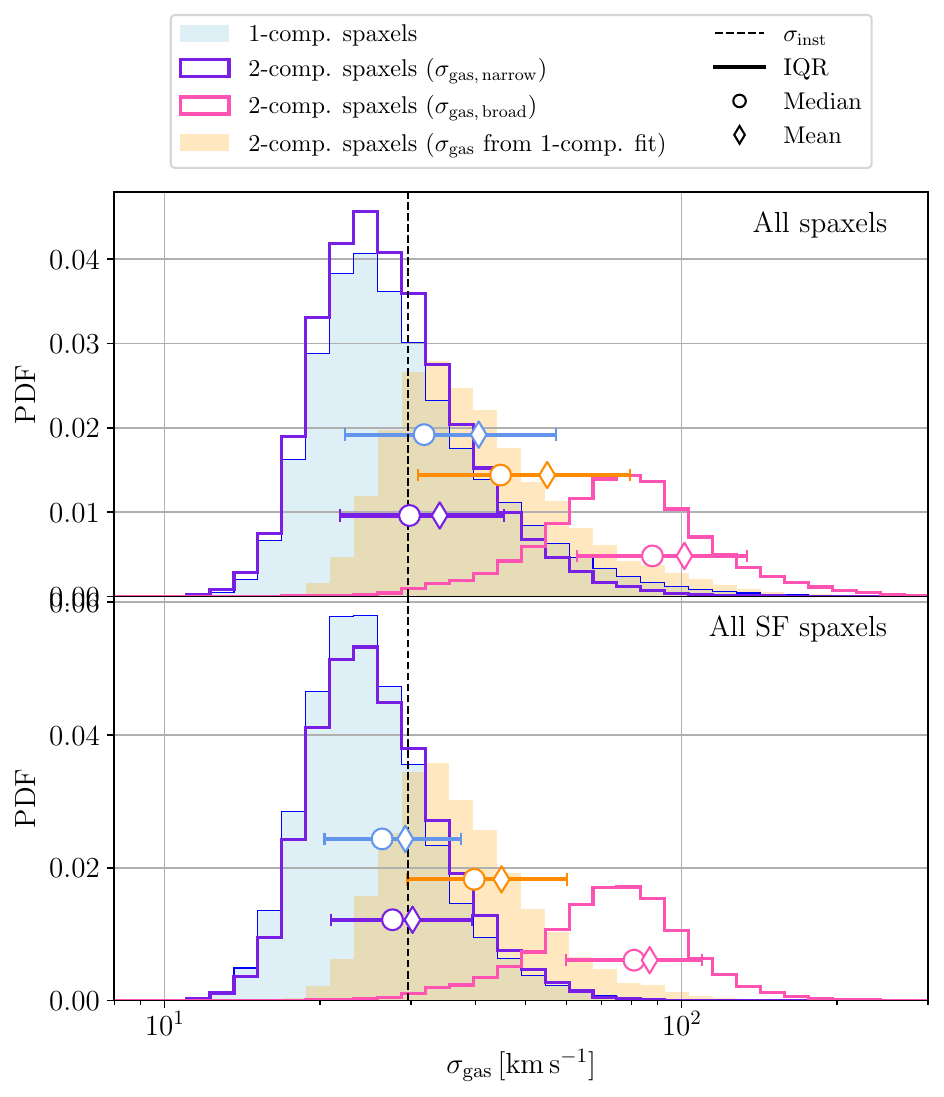}
			\caption{}
			\label{fig: observational biases: all and SF spaxels}
		\end{subfigure}
		\begin{subfigure}[b]{1.0\linewidth}
			\centering
			\includegraphics[width=\linewidth]{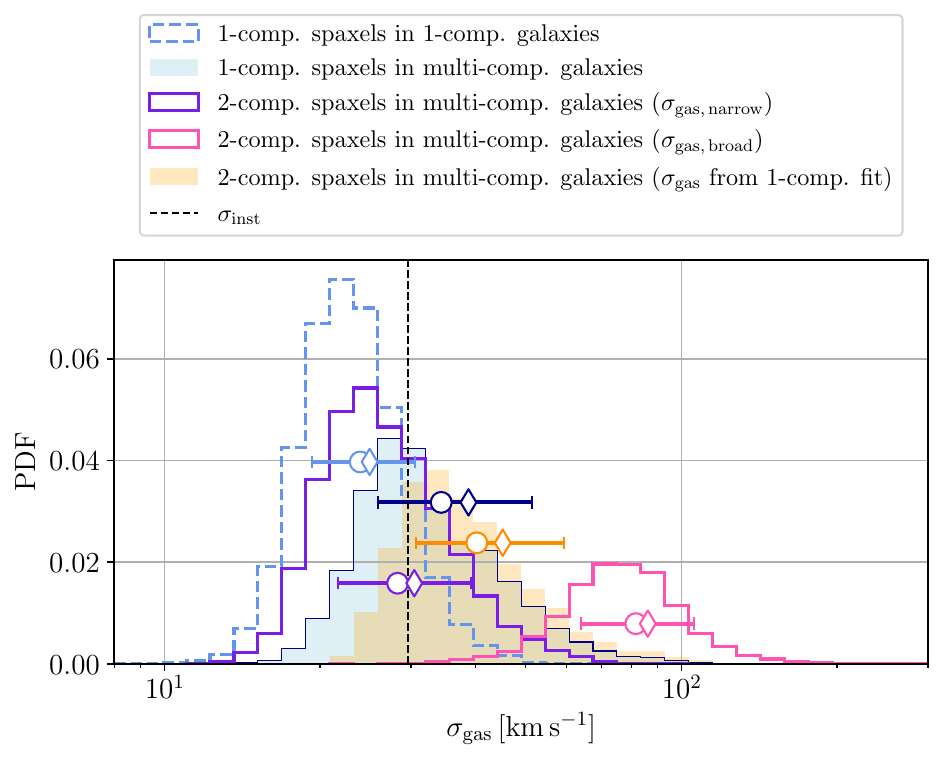}
			\caption{}
			\label{fig: observational biases: 1 comp. vs. 2 comp. galaxies}
		\end{subfigure}
		\caption{
				Velocity dispersion distributions within different subsets of spaxels. 
				Panel (a) shows normalised velocity dispersion distributions for 1-component spaxels (pale blue outlined histogram), the narrow component in 2-component spaxels (purple outlined histogram) and that of 1-component fits to 2-component spaxels (shaded orange histogram) for all spaxels (top panel) and for SF spaxels only (bottom panel).
				Panel (b) shows velocity dispersion distributions for subsets of SF spaxels within 1- and multi-component galaxies.
				1-component spaxels in 1- and multi-component galaxies are represented by the dashed blue and shaded pale blue histograms respectively. The narrow and broad components in 2-component spaxels in multi-component galaxies are shown in the purple and pink outlined histograms respectively, and the shaded orange histogram represents the 1-component fit to the 2-component spaxels. 
				Corresponding medians and means are indicated by the circles and diamonds respectively, and IQRs are shown by the error bars; values are given in Table~\ref{tab: velocity dispersion statistics}. 
				The SAMI instrumental resolution is indicated by the black dashed line. 	
		}
	\label{fig: velocity dispersion distributions}
	\end{figure}

	First, we consider all spaxels in the SAMI sample. In the top panel of Fig.~\ref{fig: observational biases: all and SF spaxels}, we show \sigmagas\ distributions for 1-component spaxels (shaded blue), \sigmanarrow\ and \sigmabroad\ of the 2-component spaxels (open purple and open pink respectively), and of \sigmagas\ from the 1-component fits to the 2-component spaxels (shaded orange).
	The corresponding means, standard deviations, medians and interquartile ranges (IQRs) are reported in Table~\ref{tab: velocity dispersion statistics}. 
	If there were a significant population of 2-component spaxels contaminating the sample of 1-component spaxels, then the blue histogram would be skewed towards values similar to the orange histogram. 
	This is indeed the case: the IQR of the 1-component spaxels is skewed to larger values of \sigmagas\ compared to the narrow component in 2-component spaxels. 
	
	However, as shown in the bottom panel of Fig.~\ref{fig: observational biases: all and SF spaxels}, among the SF spaxels, which are the focus of this work, the distribution in \sigmagas\ of the 1-component spaxels is similar to that of \sigmanarrow\ in the 2-component spaxels, whereas the distribution in \sigmagas\ of the 1-component fits to the 2-component spaxels is markedly offset by approximately 10\,\kms. The purple histogram is in fact skewed towards higher values of \sigmagas\ than the blue histogram, and the two distributions are statistically significantly different according to the KS and AD tests with $p$-values $\ll 0.05$ for both tests. This would be consistent with there being only minimal contamination by 2-component spaxels amongst the SF 1-component spaxels.
	We also note that among both the 1-component SF spaxels and the narrow component of the 2-component SF spaxels, the average velocity dispersion $\sigmagas \approx 30 \pm 10\,\kms$, which is similar to that of the ``dynamically cold'' ionised gas component associated with SF regions identified by \citet{Law2021}; meanwhile, the typical velocity dispersions of the broad component in 2-component SF spaxels is similar to that of their ``dynamically hot'' component. 
	
	We now consider how contamination of misclassified 2-component spaxels may affect our galaxy classification (Section~\ref{sec: Defining 1- and multi-component galaxies}). In Fig.~\ref{fig: observational biases: 1 comp. vs. 2 comp. galaxies}, we compare the \sigmagas\ of 1-component spaxels in 1-component galaxies (open dashed blue histogram) to \sigmanarrow\ in the 2-component spaxels in multi-component galaxies (open purple).
	In comparison with Fig.~\ref{fig: observational biases: all and SF spaxels}, there is a larger difference between the two distributions, such that the purple histogram is even more skewed towards higher values.
	This is consistent with our hypothesis that there is only minimal contamination of misclassified 2-component spaxels within the 1-comp. galaxies.
	Meanwhile, the 1-component spaxels in multi-component galaxies (shaded blue histogram) are much broader than the narrow component in the 2-component spaxels, and have a distribution skewed towards that of the 1-component fits to the 2-component spaxels, suggesting that some 1-component spaxels in the 2-component galaxies may represent misclassified 2-component spaxels. 
	
	From these two tests, we conclude that our analysis is unlikely to be affected by misclassified multi-component spaxels. Although our multi-component galaxies may contain significant numbers of broad 1-component spaxels that may be misclassified multi-component spaxels, this effect is probably minimal in the 1-component galaxies, indicating that our galaxy classification scheme is robust against contamination of misclassified multi-component spaxels.

	\begin{table*}
		\centering
		\caption{Means, standard deviations, and percentiles for the velocity dispersion distributions shown in Fig.~\ref{fig: velocity dispersion distributions}. 3-component spaxels are not included due to their rarity compared to 2-component spaxels. All values are in units of \kms.}
		\vspace{0.5cm}
		\begin{tabular}{lccccc}
			\toprule
			\textbf{Quantity} & \textbf{Mean} & \textbf{Std. dev.} & \textbf{16th percentile} & \textbf{Median} & \textbf{84th percentile} \\
			\midrule
			\multicolumn{6}{c}{\textbf{All spaxels}}\\
			\midrule
			$\sigma$ of 1-comp. spaxels & 40.54 & 27.76 & 22.31 & 31.79 & 57.14 \\
			\sigmanarrow\ of 2-comp. spaxels & 34.08 & 16.22 & 21.82 & 29.79 & 45.33 \\
			\sigmabroad\ of 2-comp. spaxels & 101.34 & 61.91 & 62.93 & 87.90 & 134.04 \\
			$\sigma$ of 1-comp. fit to 2-comp. spaxels & 55.00 & 32.14 & 30.94 & 44.69 & 79.60 \\
			\midrule
			\multicolumn{6}{c}{\textbf{SF spaxels}}\\
			\midrule
			$\sigma$ of 1-comp. spaxels & 29.24 & 11.38 & 20.40 & 26.37 & 37.45 \\
			\sigmanarrow\ of 2-comp. spaxels & 30.19 & 11.03 & 20.96 & 27.61 & 39.32 \\
			\sigmabroad\ of 2-comp. spaxels & 86.81 & 42.53 & 59.83 & 80.98 & 109.49 \\
			$\sigma$ of 1-comp. fit to 2-comp. spaxels & 44.88 & 18.79 & 29.52 & 39.74 & 60.07 \\
			\midrule
			\multicolumn{6}{c}{\textbf{SF spaxels in 1-component galaxies}}\\
			\midrule
			$\sigma$ of 1-comp. spaxels & 24.94 & 6.19 & 19.30 & 23.92 & 30.53 \\
			\midrule
			\multicolumn{6}{c}{\textbf{SF spaxels in multi-component galaxies}}\\
			\midrule
			$\sigma$ of 1-comp. spaxels & 38.74 & 15.61 & 25.87 & 34.31 & 51.37 \\
			\sigmanarrow\ of 2-comp. spaxels & 30.43 & 9.82 & 21.65 & 28.26 & 39.23 \\
			\sigmabroad\ of 2-comp. spaxels & 86.10 & 27.03 & 63.81 & 81.63 & 105.68 \\
			$\sigma$ of 1-comp. fit to 2-comp. spaxels & 45.12 & 17.55 & 30.62 & 40.20 & 59.34 \\
			\bottomrule
		\end{tabular}
		\label{tab: velocity dispersion statistics}
	\end{table*}

\subsubsection{Spectral resolution limitations}\label{subsubsec: Spectral resolution limitations}
In general, individual emission line components cannot be resolved by a spectrograph with resolution \sigmainst{} if the difference between the LoS velocities of the broad and narrow components $v_{\rm broad} - v_{\rm narrow} := \dv \lesssim \sigmainst$ and the difference in the velocity dispersions $\sigmabroad - \sigmanarrow := \dsigma \lesssim \sigmainst$. 

\begin{figure}
	\centering
	\includegraphics[width=0.7\linewidth]{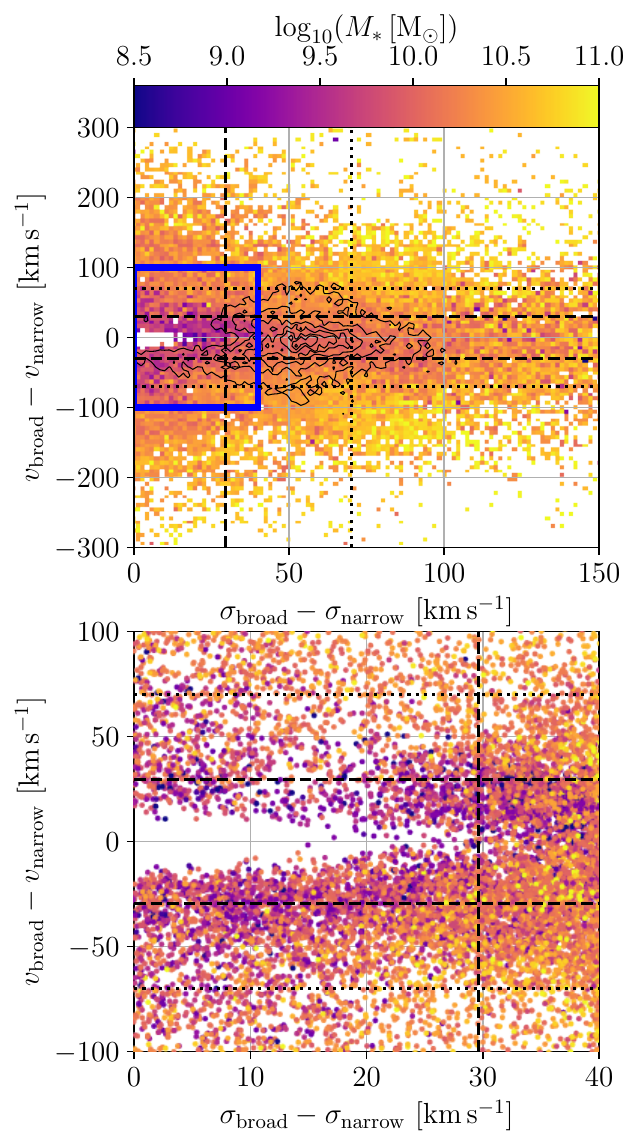}\\
	\caption{
		\dv{} vs. \dsigma{} for individual spaxels with two components, coloured by the stellar mass of their host galaxy. In the upper panel, the data is represented as a 2D histogram, in which the colour of each cell represents the median \mstar{} of all spaxels falling within that cell, and the black contours show the number distribution of spaxels on a log scale. The lower panel shows an inset view of the left-hand plot (indicated by the blue rectangle), where spaxels are shown individually as circles. The dashed lines represent the spectral resolution of SAMI (29.6\kms) and the dotted lines represent the MaNGA spectral resolution (approximately 70\,\kms), indicating that MaNGA would be unable to spectrally resolve approximately 53\,per\,cent of the multi-component spaxels identified in this work.
	}
	\label{fig: squid plots}
\end{figure}

The inability of SAMI to detect multiple emission line components closely spaced in both velocity and velocity dispersion is demonstrated in Fig.~\ref{fig: squid plots}, where we plot all 2-component spaxels in the \dv{}--\dsigma{} plane. 
Again, we do not consider 3-component spaxels here due to their rarity in comparison with 2-component spaxels.
There is a ``hole'' roughly defined by $\dsigma \lesssim \sigmainst$ and $|\dv| \lesssim \sigmainst$, at which point multiple emission line components can no longer be distinguished at the spectral resolution of SAMI. 
If $\dsigma \lesssim \sigmainst$ but $|\dv| \gg \sigmainst$, i.e. the two components have similar widths, but are sufficiently separated in LoS velocity, then the two components can be distinguished. Conversely, if $|\dv| < \sigmainst$ but $\dsigma \gg \sigmainst$, i.e. the two components have similar LoS velocities but have substantially different widths, then the two components can be distinguished as well. 
The spaxels lying closest to the ``forbidden'' region of the parameter space tend to reside in less massive galaxies; we therefore conclude that with SAMI we may be unable to reliably detect DIG/thick ionised gas disks using this method in low-mass systems, as discussed in Section~\ref{subsec: Physical interpretation}. 

In Fig.~\ref{fig: squid plots} we also indicate the spectral resolution of MaNGA~\citep[approximately 70\,\kms; e.g.,][]{Law2022} by the dotted lines. As shown by the black contours, which trace the distribution of 2-component spaxels in the \dv--\dsigma{} plane, approximately 53\,per\,cent of 2-component spaxels that can be detected with SAMI have $\dv < \sigma_{\rm inst,\,MaNGA}$, indicating that typical broad emission line components in local galaxies cannot be resolved with MaNGA using the standard spectral decomposition technique employed here. 
This demonstrates the importance of high spectral resolution when conducting detailed studies of gas kinematics.

\subsubsection{Inclination effects}\label{subsubsec: Inclincation effects}

As discussed in Section~\ref{sec: Defining 1- and multi-component galaxies} and as shown in Fig.~\ref{fig: sample histograms (observational)}, multi-component galaxies are under-represented at near-face-on and near-edge-on inclinations. Additionally, at a fixed local $\Sigma_{\rm SFR}$, spaxels residing in galaxies at higher inclinations are less likely to exhibit multiple components (Fig.~\ref{fig: f vs. local SFR surface density (binned by inclination)}).
There are several possible explanations for this.

At higher inclinations, extraplanar gas or winds above and below the disk plane may not share a LoS with gas in the thin disk, and would therefore be visible as a single, but broad, component.
In a subset of edge-on SAMI galaxies, \citet{Ho2016a} identified wind-dominated and DIG-dominated systems by assuming that winds should exhibit higher velocity dispersions and greater asymmetry in the velocity field beyond the disk plane compared to DIG. 
Of their 15 wind-dominated systems, two are present in our 1-component sample, and two of our multi-component galaxies are found.
Meanwhile, their sample of DIG-dominated edge-on systems contains 7 of our 1-component galaxies and no multi-component galaxies. 
This suggests that the presence of multiple emission line components may be an unreliable method for detecting winds and DIG in edge-on systems. 
However, their sample of edge-on galaxies was chosen to have axis ratios $b / a < 0.26$, corresponding to only 23 galaxies in our parent sample of 655 galaxies, indicating that our spectral decomposition technique is unlikely to be missing a large number of wind- and DIG-dominated galaxies in our sample due to edge-on inclination effects. Of the 18 high-$\Sigma_{\rm SFR}$ 1-component galaxies discussed in Section~\ref{subsubsec: high-SFR surface density 1-component galaxies}, only two have $b / a < 0.26$, indicating that the lack of observed complex line profiles in these objects is unlikely to result from inclination effects.

\begin{figure}
	\centering
	\includegraphics[width=0.8\linewidth]{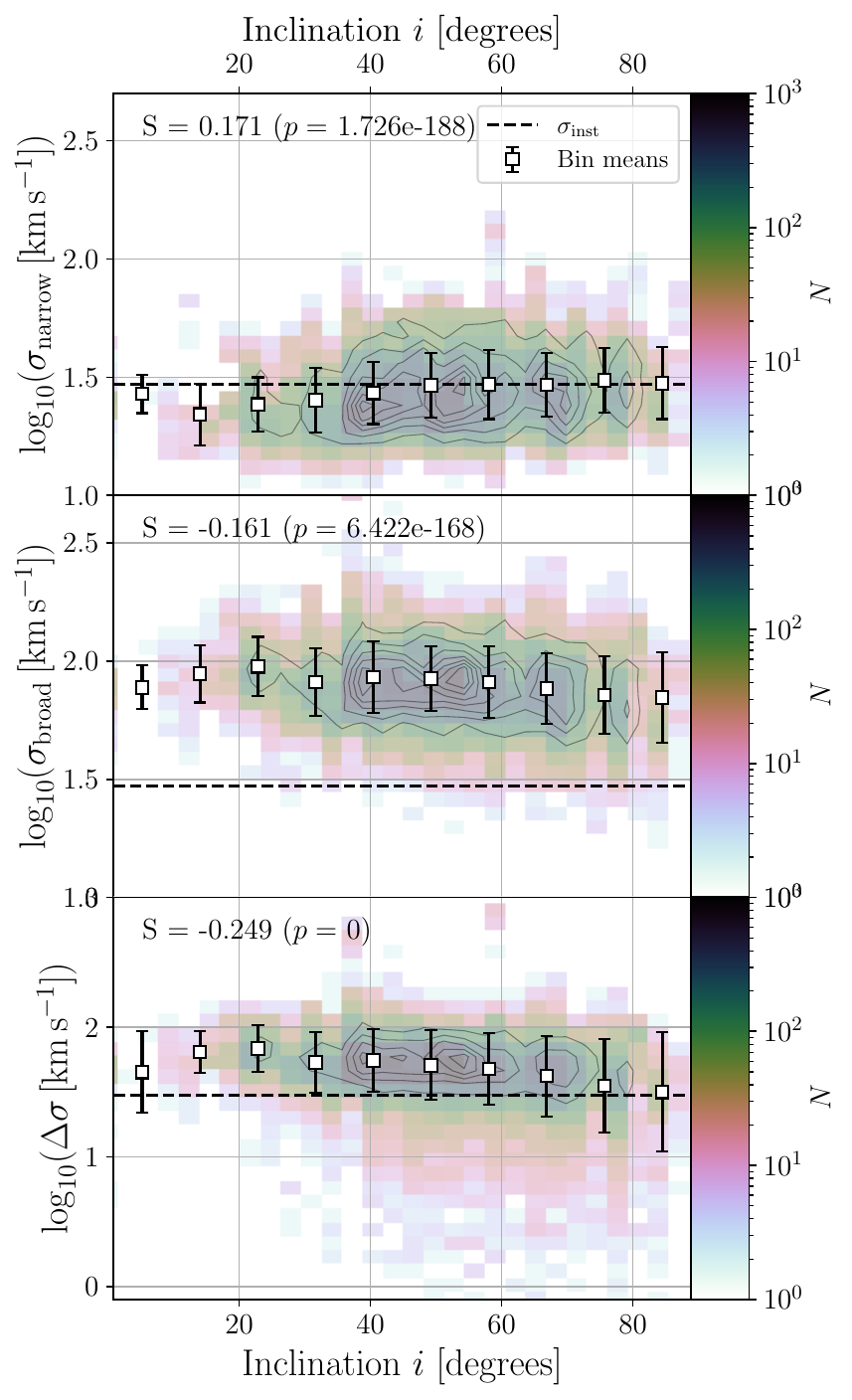}
	\caption{Distribution of all 2-component spaxels in \sigmanarrow, \sigmabroad{} and \dsigma{} as a function of inclination. The white squares show median values and the error bars represent the standard deviation in different inclination bins. The horizontal dashed lines represent the SAMI instrumental resolution of 29.6 \kms. The Spearman's rank-order correlation coefficient and accompanying $p$-value is shown in the upper right-hand corner.}
	\label{fig: velocity dispersion vs. inclination}
\end{figure}

The lack of high-inclination multi-component galaxies may also be due to spectral resolution limitations.
In Fig.~\ref{fig: velocity dispersion vs. inclination} we plot the velocity dispersions of the narrow and broad components (\sigmanarrow{} and \sigmabroad{} respectively), plus \dsigma{}, as a function of inclination. 
\sigmanarrow{} is weakly correlated with inclination, most likely due to the artificial broadening of emission lines in more highly-inclined systems due to rotation, whereas \sigmabroad{} is anticorrelated.
The observed trends may arise if the vertical component of the velocity dispersion $\sigma_{zz}$ is larger than the azimuthal or radial components $\sigma_{\phi\phi}$ and $\sigma_{rr}$. Such anisotropy may be expected in the presence of galactic winds~\citep{Marinacci2010a} where material is preferentially launched out of the disk plane. However, this explanation contradicts the findings of \citet{Law2022}, who found that $\sigma_{zz} < \sigma_{\phi\phi}, \sigma_{rr}$ in a sample of local star-forming galaxies, although their analysis is based upon the \ha{} velocity dispersion measured from single Gaussian fits to lower-resolution MaNGA data.

Rather, the anticorrelation between \sigmabroad{} and inclination most likely originates from observational effects. 
Galactic fountains may lead to the formation of a thick ionised gas disk which has a negative vertical velocity gradient resulting in a slower circular velocity than the thin disk~\citep{Swaters1997,Levy2018,Levy2019,Marasco2019}.
In this configuration, the LoS velocity separation between the thin and thick disks increases as a function of inclination, whilst in face-on systems $\dv \sim 0\,\kms$. 
Referring to Fig.~\ref{fig: squid plots}, SAMI is unable to resolve multiple emission line components when $\dv \lesssim \sigmainst$ and $\dsigma \lesssim \sigmainst$. Therefore, in face-on systems, we may only spectrally resolve broad emission line components from thick disks if $\dsigma > \sigmainst$, whereas in highly inclined systems, spaxels with $\dsigma < \sigmainst$ may be resolved as long as $\dv > \sigmainst$. As a result, at low inclinations, we can only detect thick disks in which $\dsigma > \sigmainst$, whereas at higher inclinations we may detect thick disks with a wider range of velocity dispersions. Our inability to spectrally resolve face-on thick disks may therefore explain the lack of spaxels with $\dsigma < \sigmainst$ at inclinations $\lesssim 20^\circ$.

In wind-dominated systems, we can expect $\dv \gg 0\,\kms$ in low-inclination systems, meaning that in principle it would be possible to resolve a broad component with $\dsigma < \sigmainst$. However, winds typically have $\dsigma \gg \sigmainst$ due to large turbulent motions and outflow velocities; this is therefore consistent with the lack of spaxels with $\dsigma < \sigmainst$ at low inclinations.

Finally, it is also possible that multi-component galaxies are intrinsically ``puffier'' than their 1-component counterparts, biasing inclination measurements towards lower values due to our assumption of a fixed disk thickness of $q_0 = 0.2$.


\section{Conclusions}

We applied a spectral decomposition technique to distinguish line emission from the star-forming thin disk and winds and/or thick disks in star-forming galaxies from the SAMI Galaxy Survey. Our conclusions are as follows:

\begin{enumerate}
	\item The presence of complex emission line profiles in star-forming galaxies is strongly correlated with the global $\Sigma_{\rm SFR}$ of the host galaxy, even when controlling for sample biases in stellar mass and SFR, and observational effects such as inclination, angular scale and A/N. This correlation is consistent with previous studies that have found a strong connection between $\Sigma_{\rm SFR}$ and both thick ionised gas disks and winds. In fact, virtually all galaxies in our sample that contain a significant fraction of spaxels with complex emission line profiles have $\Sigma_{\rm SFR} \gtrsim 10^{-2} \,\msol \, \yr^{-1} \, \kpc^{-2}$, which is remarkably similar to the value found by \citet{Roberts-Borsani2020a}. 
	\item Whilst \textit{global} $\Sigma_{\rm SFR}$, measured via \sfrdens{\re}, is the quantity that is most strongly correlated with the fraction $F_{\re}$ of multi-component spaxels within individual galaxies within $1\re$, the fraction $f_n$ of individual spaxels containing multiple components is also strongly dependent on the local $\Sigma_{\rm SFR}$ measured within individual spaxels. We also find that, at a fixed local $\Sigma_{\rm SFR}$, spaxels in galaxies with higher global $\Sigma_{\rm SFR}$ are more likely to exhibit multiple components. This illustrates the importance of considering global galaxy properties when conducting studies of individual spaxels in IFS surveys.
	\item There is a strong residual correlation between \sfrdens{\re} and $F_{\re}$ when controlling for Gini coefficient $G$, suggesting that global $\Sigma_{\rm SFR}$ is more important than clumpiness in dictating the presence of widespread complex emission line profiles within a galaxy.
	\item The strong correlation between $\Sigma_{\rm SFR}$ and the presence of complex emission line profiles is consistent with the findings of previous studies that have employed different techniques to identify winds and thick ionised gas disks. This indicates that the spectral decomposition technique is an effective means of identifying these phenomena in local galaxies.
	\item SAMI is currently the only IFS survey with sufficient spectral resolution to resolve multiple emission line components in local galaxies using the spectral decomposition technique employed here. However, SAMI is likely unable to detect broad emission line components in low-mass galaxies, where the LoS velocity and velocity dispersion offsets between the narrow and broad components are smaller than the instrumental resolution. Hector~\citep{Bryant2020}, the successor to SAMI, will have a comparable spectral resolution, enabling similar studies to be carried out on a much larger sample of over 10,000 galaxies. 
\end{enumerate}

\section*{Acknowledgements}

H. R. M. Z. thanks K. Barger, S. Barsanti, D. Fisher, M. Krumholz, S. Vaughan, A. Vijayan, and E. Wisnioski for valuable discussions.

The SAMI Galaxy Survey is based on observations made at the Anglo-Australian Telescope. The Sydney-AAO Multi-object Integral field spectrograph (SAMI) was developed jointly by the University of Sydney and the Australian Astronomical Observatory. The SAMI input catalogue is based on data taken from the Sloan Digital Sky Survey, the GAMA Survey and the VST ATLAS Survey. The SAMI Galaxy Survey is supported by the Australian Research Council Centre of Excellence for All Sky Astrophysics in 3 Dimensions (ASTRO3D), through project number CE170100013, the Australian Research Council Centre of Excellence for All-sky Astrophysics (CAASTRO), through project number CE110001020, and other participating institutions. The SAMI Galaxy Survey website is \url{http://sami-survey.org/}.

This research was supported by the Australian Research Council Centre of Excellence for All Sky Astrophysics in 3 Dimensions (ASTRO 3D), through project number CE170100013. 
S. O. acknowledges support from the NRF grant funded by the Korea government (MSIT) (No. 2020R1A2C3003769 and No. RS-2023-00214057).
F. D. E. acknowledges funding through the ERC Advanced grant 695671 ``QUENCH'' and support by the Science and Technology Facilities Council (STFC).
L. C. acknowledges support from the Australian Research Council Discovery Project and Future Fellowship funding schemes (DP210100337, FT180100066).

This work made extensive use of the \textsc{python} packages \mbox{\textsc{Astropy}}\footnote{\url{http://www.astropy.org}}\citep{Astropy2013,Astropy2018,Astropy2022}, \mbox{\textsc{Extinction}}\footnote{\url{https://extinction.readthedocs.io/en/latest/}}, \mbox{\textsc{NumPy}}\footnote{\url{https://numpy.org/}}\citep{Harris2020}, \mbox{\textsc{SciPy}}\footnote{\url{https://scipy.org/}}\citep{Virtanen2020} \new{and  \mbox{\textsc{pingouin}}\footnote{\url{https://pingouin-stats.org/}}\citep{Vallat2018}}.
The figures in this paper were produced using \mbox{\textsc{Matplotlib}}\footnote{\url{https://matplotlib.org/stable/index.html\#}}\citep{Hunter2007}.

The majority of this research was carried out on the traditional lands of the Ngunnawal and Ngambri people, and the observations presented in this work were gathered at Siding Spring Observatory, located on the traditional lands of the Gamilaraay/Kamilaroi people.

\section*{Data Availability}

The data used in this paper is from the SAMI Galaxy Survey Data Release 3 which is available at \url{https://docs.datacentral.org.au/sami/}. \textsc{spaxelsleuth}, the \textsc{python} package used to carry out all analysis in this work, is available at \url{https://github.com/hzovaro/spaxelsleuth}. 



\bibliographystyle{mnras}
\bibliography{bibliography.bib} 




\appendix
\section{Example \lzifu{} fits}\label{app: Example LZIFU fits}
\begin{figure*}
	\centering
	\begin{subfigure}[t]{0.33\linewidth}
		\includegraphics[width=\linewidth]{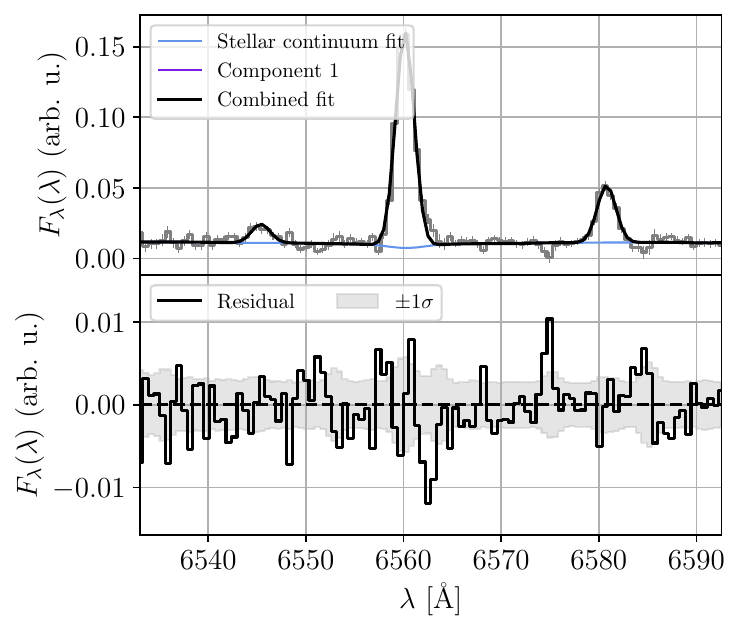}
		\caption{}
	\end{subfigure}
	\begin{subfigure}[t]{0.33\linewidth}
		\includegraphics[width=\linewidth]{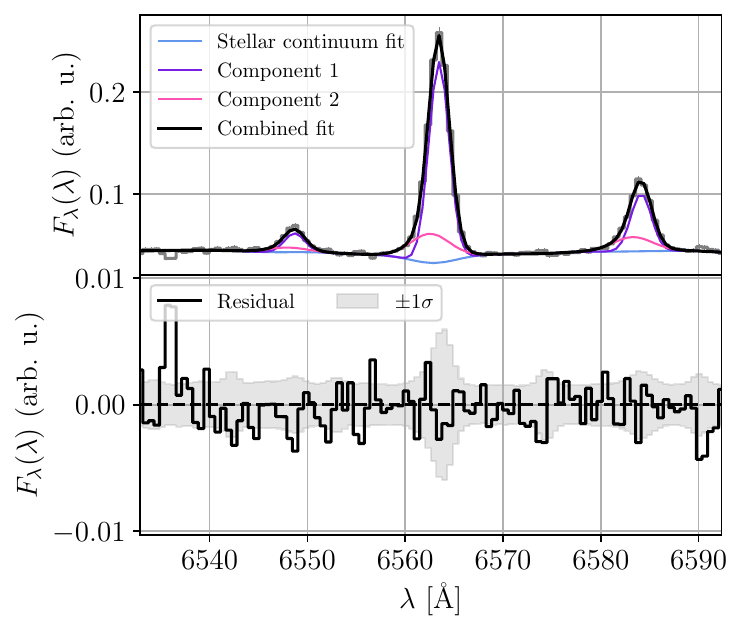}
		\caption{}
	\end{subfigure}
	\begin{subfigure}[t]{0.33\linewidth}
		\includegraphics[width=\linewidth]{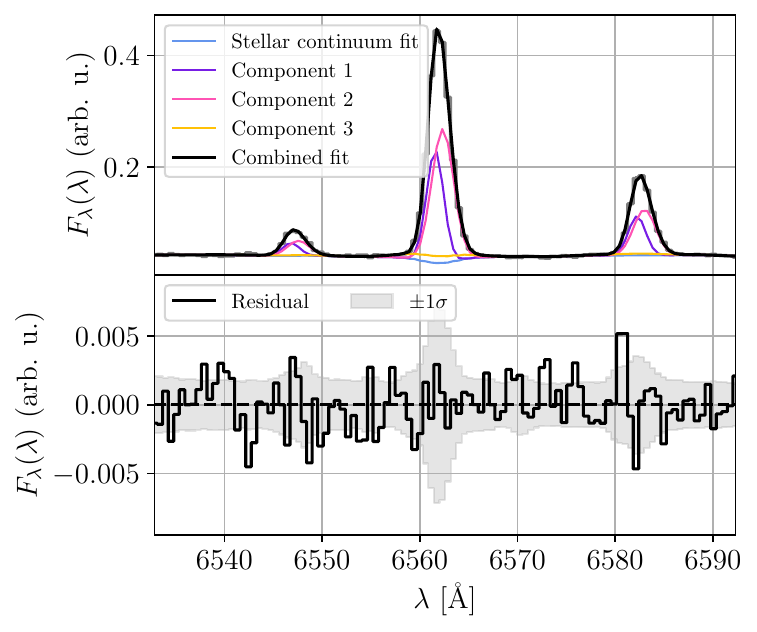}
		\caption{}
	\end{subfigure}
	\caption{
		Example \lzifu{} fits to the \ha{} and \forb{N}{ii}$\uplambda\uplambda 6548,83$ lines from randomly chosen spaxels with one (a), two (b) and three (c) components meeting the S/N and data quality cuts discussed in Section~\ref{subsec: Data quality and S/N cuts}. In each figure, the upper panel shows the data plus $1\sigma$ errors in grey, the stellar continuum fit in light blue, and individual Gaussian components for the narrow, broad and extra-broad components in purple, pink and orange respectively, plus the total fit in black. The lower panel shows the residual in black with $\pm 1\sigma$ errors indicated in grey.
	}
	\label{fig: example LZIFU fits}
\end{figure*}

Example emission line fits to a selection of randomly-chosen 1, 2 and 3-component spaxels is shown in Fig.~\ref{fig: example LZIFU fits}.

\section{Unresolved selection criteria}\label{app: Unresolved selection criteria}

As discussed in Section~\ref{sec: Defining 1- and multi-component galaxies}, we repeated our 1- and multi-component galaxy classification using aperture spectra to see whether there are any galaxies with low surface-brightness broad emission lines not detected in individual spaxels, but present in aperture spectra. 

Following \citet{Avery2021}, aperture spectra were created by summing spaxels within 1\re{} after having the LoS velocity subtracted (measured from the DR3 1-component fits to the resolved data), to minimise the effects of beam smearing on the final line widths. 1- and 2-component emission line fits were carried out using \lzifu{}~\citep{Ho2016b}.

We classified multi-component galaxies as those for which the likelihood ratio test (LRT) described in \citet{Ho2014} preferred the 2-component emission line fit to the 1-component fit at a statistical significance of $p < 0.05$; 1-component galaxies were those not meeting this criterion, but in which the 1-component fit was preferred to a stellar continuum-only fit at $p < 0.05$. 
In addition to these, we required that the spectral classification be star-forming (using the total fluxes summed over all fitted components); in all fitted components, both \ha{} and \forb{N}{ii}$\uplambda 6583Å$ must have flux S/N $>5$, flux A/N $>3$, and $\sigma$ S/N of 3 as per Section~\ref{subsec: Data quality and S/N cuts}. 
This resulted in 746 1-component galaxies and 125 multi-component galaxies. 

The unresolved scheme identified over twice as many 1-component galaxies (746 versus 266), likely due to the inclusion of galaxies with fewer than 50 spaxels with sufficient emission line S/N to be classified using the resolved data.
However, there was considerable disagreement between the resolved and unresolved classification schemes.
Of the galaxies classified as 1- and multi-component using the resolved criteria, 31 and 42 respectively were unclassified using the aperture spectra; and of those classified as 1- and multi-component using the aperture spectra, 496 and 50 were unclassified using the resolved criteria. Only 222 1-component and 68 multi-component galaxies had consistent classifications using both methods.

%
%
%
%
%

Interestingly, there were no multi-component galaxies identified by the unresolved criteria that were classified as 1-component galaxies using the spatially resolved data, suggesting that there are no galaxies with faint broad components that can only be observed in aperture spectra extracted using this technique. 
In fact, there were 20 galaxies classified as multi-component using the resolved data that were classified only as 1-component galaxies using the aperture spectra. 

There are several potential explanations for this; firstly, faint broad components that are detectable in individual spaxels may be overwhelmed by very strong narrow-component emission from other spaxels when summed to create the aperture spectra. 
Additionally, the LoS velocity subtraction was carried out using the single-Gaussian component fits in DR3, which are heavily biased towards the narrow component, as it is generally much stronger than any broad components. Whilst this step minimises the effects of beam smearing on the narrow component, the broad component may be significantly affected by beam smearing as it traces a distinct structure with separate kinematics. This may reduce its amplitude relative to the narrow component to the extent that it falls below the A/N criterion, meaning these galaxies will not be classified as multi-component galaxies. 
Lastly, any multi-component spaxels located at $r > R_e$ would not be included in the aperture spectra.
These factors demonstrate the utility of spatially resolved spectroscopy for analysing detailed gas kinematics. 
We therefore opted to use only the resolved data for this work.

\section{Additional plots and tables}\label{app: additional plots and tables}

\begin{figure*}
	\centering
	\includegraphics[width=0.8\linewidth]{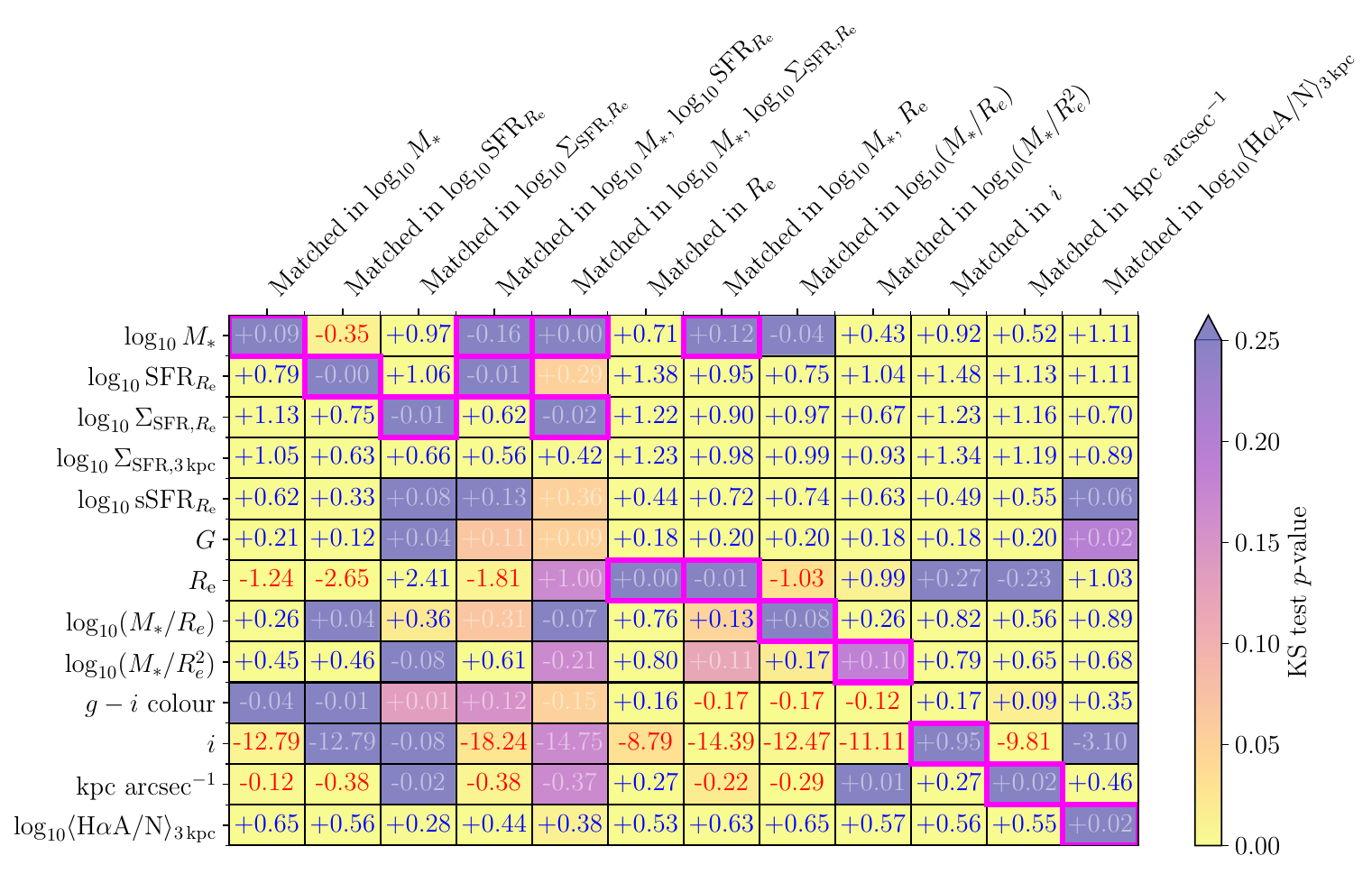}
	\caption{
		Results from the matched sample analysis in Section~\ref{sec: Matched sample analysis}. 
		Each cell compares the distributions in the \textit{row} quantity between the 1- and multi-component galaxies when matched in the \textit{column} quantity. 
		The cell colour represents the $p$-value from the two-sample KS applied to the matched samples measured in the \textit{row} quantity; in dark-coloured cells the distributions are similar, whereas pale colours indicate that the distributions are highly unlikely to be drawn from the same parent sample, i.e., that the 1- and multi-component matched samples are significantly different.
		The number in each cell represents the difference in the median values of the row quantity between the 1- and multi-component matched samples. Cells in which the $p$-value is below the significance threshold of 0.05 are indicated by coloured text, where red values indicate a \textit{lower} median value and blue values a \textit{higher} median value in the multi-component matched sample than the 1-component matched sample.
		For example, when matched in \sfr{} (2nd column), the stellar mass distributions of the 1- and multi-component galaxies (top row) are significantly different, as indicated by the bold text and pale cell colour, and the matched multi-component galaxies have a median stellar mass 0.35 dex lower than that of the 1-component galaxies. Meanwhile, the matched samples have similar distributions in $\log_{10} \left(M_* / \re\right)$ (8th row) as indicated by the dark cell colour.
	}
	\label{fig: matched sample analysis - punnet square}
\end{figure*}

Fig.~\ref{fig: matched sample analysis - punnet square} shows a ``punnet square'' plot showing the median offsets between various parameters between the 1- and multi-component samples matched in a number of quantities.

\begin{figure}
	\centering
	\includegraphics[width=\linewidth]{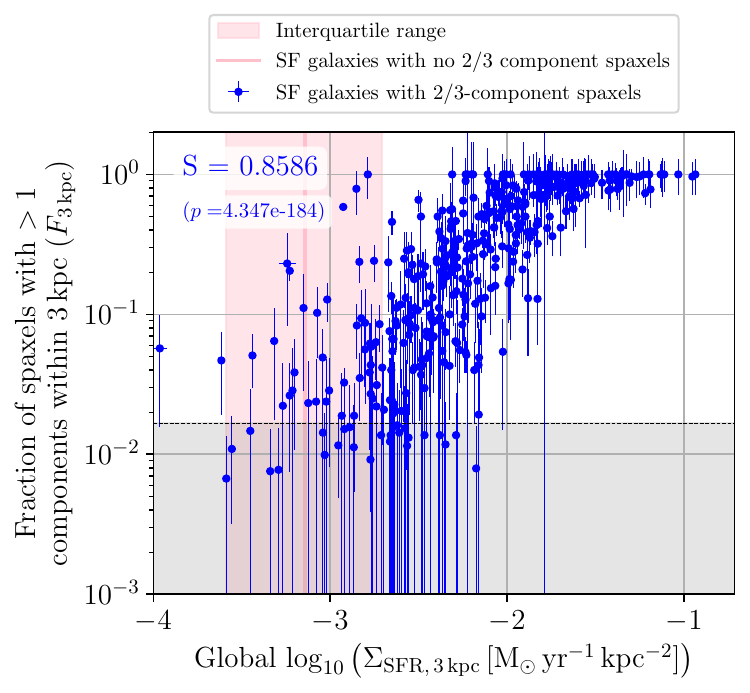}
	\caption{
		Same as Fig.~\ref{fig: F vs. global SFR surface density} with \sfrdens{\rm 3\,kpc} substituted for \sfrdens{\re}.
	}
	\label{fig: F vs. global SFR surface density (3kpc)}
\end{figure}

Fig.~\ref{fig: F vs. global SFR surface density (3kpc)} shows the counterpart of Fig.~\ref{fig: F vs. global SFR surface density} made using \sfrdens{\rm 3\,kpc} rather than \sfrdens{\re}.

\begin{figure*}
	\centering
	\begin{subfigure}[t]{0.49\linewidth}
		\includegraphics[width=\columnwidth]{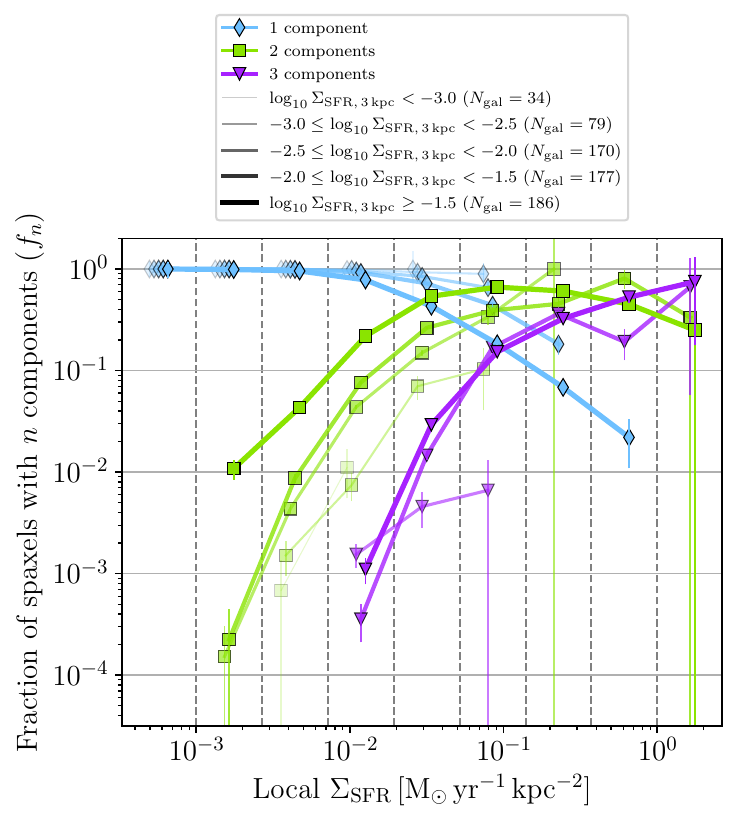}
		\caption{}
		\label{fig: f vs. local SFR surface density (binned by SFR surface density (3kpc))}
	\end{subfigure}
	\begin{subfigure}[t]{0.49\linewidth}
		\includegraphics[width=\linewidth]{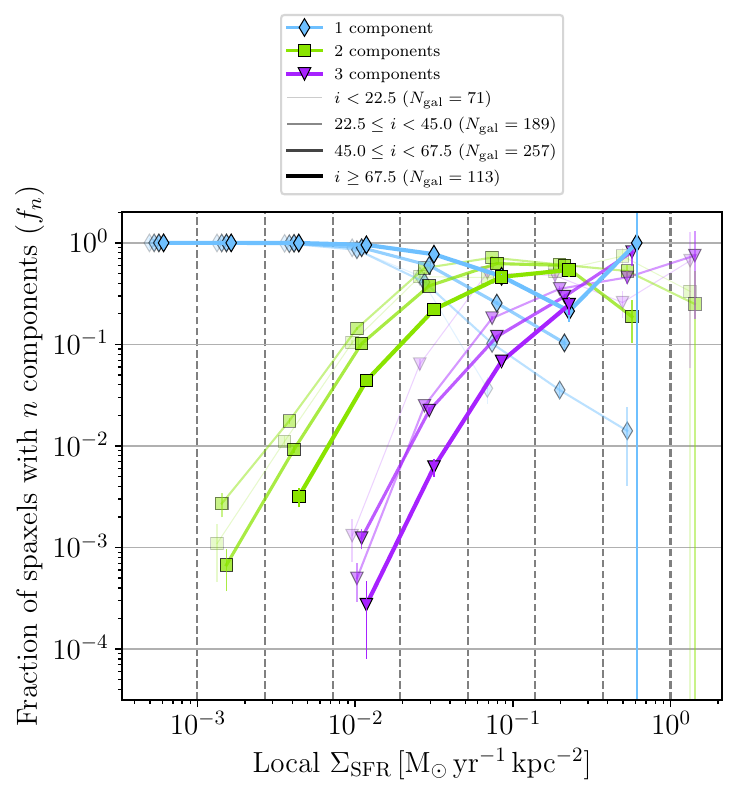}
		\caption{}
		\label{fig: f vs. local SFR surface density (binned by inclination)}
	\end{subfigure}
	\begin{subfigure}[t]{0.49\linewidth}
		\includegraphics[width=\linewidth]{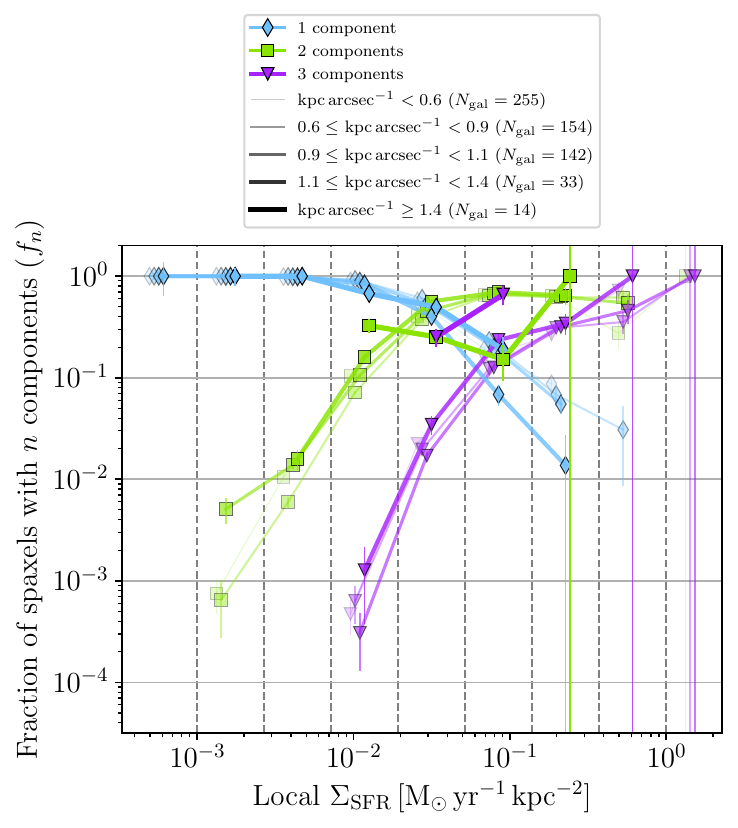}
		\caption{}
		\label{fig: f vs. local SFR surface density (binned by angular scale)}
	\end{subfigure}
	\begin{subfigure}[t]{0.49\linewidth}
		\includegraphics[width=\linewidth]{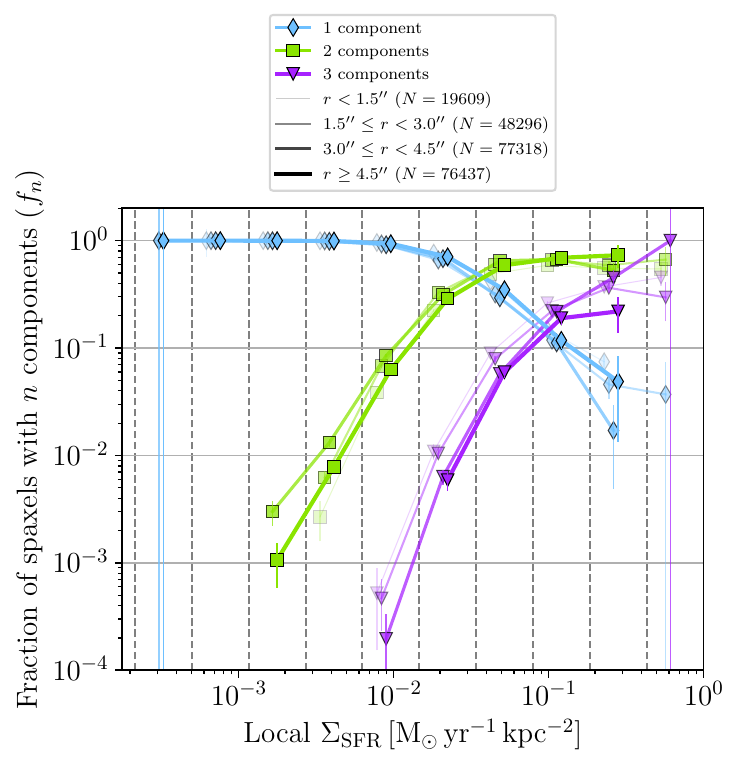}
		\caption{}
		\label{fig: f vs. local SFR surface density (binned by spaxel radius)}
	\end{subfigure}
	\caption{
		Same as Fig.~\ref{fig: f vs. local SFR surface density (binned by SFR surface density)}
		but binned by (a) \sfrdens{\rm 3\,kpc}, (b) host galaxy inclination, (c) angular scale and (d) projected spaxel radius. 
	}
	\label{fig: f vs. local SFR surface density (binned by other quantities)}
\end{figure*}

Fig.~\ref{fig: f vs. local SFR surface density (binned by other quantities)} shows $f_n$ as a function of local $\Sigma_{\rm SFR}$ binned by \sfrdens{\rm 3\,kpc}, host galaxy inclination, angular scale, and projected spaxel radius.

\bsp	
\label{lastpage}
\end{document}